\documentclass{aastex62}

\accepted{June 26, 2019}
\submitjournal{ApJ}

\shorttitle{Cyanopolyyne Chemistry around MYSOs}
\shortauthors{Taniguchi et al.}

\begin{document}

\title{Cyanopolyyne Chemistry around Massive Young Stellar Objects}

\correspondingauthor{Kotomi Taniguchi}
\email{kt8pm@virginia.edu}

\author[0000-0003-4402-6475]{Kotomi Taniguchi}
\altaffiliation{Virginia Initiative on Cosmic Origins Fellow}
\affiliation{Department of Astronomy, University of Virginia, Charlottesville, VA 22904, USA}
\affiliation{Department of Chemistry, University of Virginia, Charlottesville, VA 22903, USA}

\author[0000-0002-4649-2536]{Eric Herbst}
\affiliation{Department of Astronomy, University of Virginia, Charlottesville, VA 22904, USA}
\affiliation{Department of Chemistry, University of Virginia, Charlottesville, VA 22903, USA}

\author[0000-0003-1481-7911]{Paola Caselli}
\affiliation{Max-Planck-Institute for Extraterrestrial Physics (MPE), Giessenbachstr, 1, D-85748 Garching, Germany}

\author[0000-0002-6001-8048]{Alec Paulive}
\affiliation{Department of Chemistry, University of Virginia, Charlottesville, VA 22903, USA}

\author[0000-0002-4483-1733]{Dominique M. Maffucci}
\affiliation{Department of Chemistry, University of Virginia, Charlottesville, VA 22903, USA}

\author[0000-0003-0769-8627]{Masao Saito}
\affiliation{National Astronomical Observatory of Japan (NAOJ), Osawa, Mitaka, Tokyo 181-8588, Japan}
\affiliation{Department of Astronomical Science, School of Physical Science, SOKENDAI (The Graduate University for Advanced Studies), Osawa, Mitaka, Tokyo 181-8588, Japan}



\begin{abstract}
Recent radio astronomical observations have revealed that HC$_{5}$N, the second shortest cyanopolyyne (HC$_{2n+1}$N), is abundant around some massive young stellar objects (MYSOs), which is not predicted by classical carbon-chain chemistry.  
For example, the observed HC$_{5}$N abundance toward the G28.28$-$0.36 MYSO is higher than that in L1527, which is one of the warm carbon chain chemistry (WCCC) sources, by more than one order of magnitude \citep{2017ApJ...844...68T}.
In this paper, we present chemical simulations of hot-core models with a warm-up period using the astrochemical code Nautilus.
We find that the cyanopolyynes are formed initially  in the gas phase and accreted onto the bulk and surface of granular ice mantles during the lukewarm phase, which occurs at $25 < T < 100$ K.
In slow warm-up period models,  the peak  abundances occur as the cyanopolyynes desorb from dust grains after the temperature rises above 100 K.
The lower limits of the abundances of HC$_{5}$N, CH$_{3}$CCH, and CH$_{3}$OH observed in the G28.28$-$0.36 MYSO can be reproduced in our hot-core models, after their desorption from dust grains.
Moreover, previous observations  suggested chemical diversity in envelopes around different MYSOs.  
We discuss possible interpretations of relationships between stages of the star-formation process and such chemical diversity, such as the different warm-up timescales.   
This timescale depends not only on the mass of central stars but also on the relationship between the size of warm regions and their infall velocity.
\end{abstract}



\section{Introduction} \label{sec:intro}

Knowledge of the chemical composition allows us to use molecules as powerful diagnostic tools of the physical conditions and dynamical evolution of star-forming regions \citep[e.g.,][]{2012A&ARv..20...56C,2018IAUS..332....3V}.
Recent improvements in efficiencies of radio telescopes enable the achievement of  high sensitivities with high angular resolution within reasonable observing times.
Owing to such developments, we can detect molecules with abundances significantly lower than previously detected species \citep[e.g.,][]{2018Sci...359..202M}, including isotopologues \citep{2016ApJ...817..147T,2017PASJ...69L...7T,2018MNRAS.474.5068B}.
The discovery of these molecules in various evolutionary stages raises new challenges in the fields of chemical network simulation and laboratory experiment.

Unsaturated carbon-chain species, such as C$_{2n}$H, CCS, and the cyanopolyynes (HC$_{2n+1}$N), are unique molecules in the interstellar medium.
For a long time, they have been thought to be abundant in young starless cores and decrease in abundance in later stages of low-mass star formation \citep{1998ApJ...506..743B,2009ApJ...699..585H,1992ApJ...392..551S}.
The precursors of carbon-chain molecules in starless cores are mainly ionic carbon (C$^{+}$) and atomic carbon (C).
Carbon-chain molecules are mainly formed via gas-phase ion-molecule reactions and neutral-neutral reactions, the reaction rate coefficients of which can be large even at low temperatures such as $T \simeq 10$ K.

In contrast to the above classical picture,  low-mass protostellar cores rich in carbon-chain species have been found in sources such as L1527 in Taurus \citep{2008ApJ...672..371S} and IRAS15398--3359 in Lupus \citep{2009ApJ...697..769S}. On the other hand,
 saturated complex organic molecules (COMs) have been found to be abundant around protostars \citep{2009ARA&A..47..427H} in  so-called hot cores and hot corinos in high- and low-mass star-forming regions, respectively.
Therefore, the discovery of such low-mass protostellar cores   rich in carbon-chain species was surprising initially. 
The chemistry that produces these species was named ``warm carbon chain chemistry'' \citep[WCCC;][]{2013ChRv..113.8981S}.
In WCCC, the formation of carbon-chain molecules starts with the desorption of CH$_{4}$ from dust grains at a temperature of $\sim25$ K.
The reaction between CH$_{4}$ and C$^{+}$ and subsequent gas-phase reactions then lead to the formation of carbon chains \citep{2008ApJ...681.1385H}.
The different timescale of prestellar collapse was proposed as an origin of the difference between WCCC and hot corino chemistry \citep{2013ChRv..113.8981S}; the short and long starless core phases lead to WCCC and hot corino sources, respectively.
On the other hand, \citet{2016A&A...592L..11S} suggested that different illumination by the interstellar radiation field around dense cloud cores could produce such chemical differentiation, based on their observations toward the L1544 prestellar core.
Hence, the origin of the chemical diversity found in low-mass star-forming regions is still controversial.

Observational studies of carbon-chain molecules in high-mass star-forming regions tend to be less detailed  than those of low-mass star-forming regions.
Nevertheless, cyanopolyynes have been detected in some famous high-mass star-forming regions such as Sgr B2 \citep{2013A&A...559A..47B,2017A&A...604A..60B} and the Orion region, Orion KL \citep{2013A&A...559A..51E} and OMC-2 FIR4 \citep{2017A&A...605A..57F}.
Survey observations of HC$_{3}$N and HC$_{5}$N have been conducted toward high-mass starless cores (HMSCs) and high-mass protostellar objects (HMPOs) using the Nobeyama 45-m radio telescope \citep{2018ApJ...854..133T,2019ApJ...872..154T}.
This group detected HC$_{3}$N in almost all of the target sources and detected HC$_{5}$N in  half of the HMPOs.
Indeed, HC$_{3}$N seems to be ubiquitous around HMPOs.
\citet{2017ApJ...844...68T} found that HC$_{5}$N is abundant in envelopes around three massive young stellar objects (MYSOs) associated with the 6.7 GHz methanol masers using the Green Bank 100-m and Nobeyama 45-m telescopes.
The HC$_{5}$N abundance in G28.28$-$0.36, which is one of the target MYSOs, is higher than that in the Class 0 protostar L1527 by a factor of 20.
In addition, \citet{2018ApJ...866..150T} suggested chemical diversity around MYSOs as a reason for the varying cyanopolyyne abundances; organic-poor MYSOs are surrounded by a cyanopolyyne-rich lukewarm envelope, while organic-rich MYSOs, namely hot cores, are surrounded by a CH$_{3}$OH-rich lukewarm envelope.

In this paper, we report an investigation of  the cyanopolyyne chemistry using  hot-core models with a warm-up period, motivated by the particularly high abundance of HC$_{5}$N in the G28.28$-$0.36 MYSO.
Cyanopolyynes are relatively stable species compared with other carbon-chain species and can survive in high temperature regions \citep{2011ApJ...743..182H}.
In fact, the vibrationally excited lines of HC$_{3}$N and the extremely high-$J$ rotational lines of HC$_{5}$N have been detected in Sgr B2(N) and Orion KL \citep{2017A&A...604A..60B,2013A&A...559A..51E}.
In addition, HC$_{3}$N and HC$_{5}$N were detected in a candidate position of the molecular outflow around  G28.28$-$0.36  \citep{2018ApJ...866...32T}.
Therefore, there is a large possibility that cyanopolyynes exist in hot core regions, where the temperature is higher than in WCCC sources, and where formation mechanisms of cyanopolyynes may differ in the two types of regions.
Furthermore, the formation paths of cyanopolyynes  have been thought to consist of possible steps in the synthesis of prebiotic molecules \citep{2017A&A...597A..40J}, so it is also important to reveal their formation paths for science related to the origins of life.

In Section \ref{sec:model}, we describe our models, with results found in Section \ref{sec:res}.
We compare models, and the model results with the observational results around MYSOs \citep{2017ApJ...844...68T,2018ApJ...866..150T} in Sections \ref{sec:dis1} and \ref{sec:dis2}, respectively. 
Effects of the cosmic-ray ionization rate on the HC$_{5}$N abundance are investigated in Section \ref{sec:dis5}.
The chemistry of cyanopolyynes is compared with that of other carbon-chain species during the warm-up period in Section \ref{sec:dis3}.
Possible origins of the chemical diversity around MYSOs are discussed in Section \ref{sec:dis4}. 
Finally, our conclusions are presented in Section \ref{sec:con}.

\section{Models} \label{sec:model}

In this study, we used the gas-grain Nautilus code \citep{2016MNRAS.459.3756R} supplemented by preliminary reactions involving irradiation \citep{2018PCCP...20.5359S} to model the free-fall collapse and warm-up periods of hot core evolution \citep{2006A&A...457..927G}.
We ran both the three-phase model, in which the chemistry of the grain surface and the bulk ice are distinguished, and the two-phase model, in which the grain surface and bulk ice are not distinguished \citep{2016MNRAS.459.3756R}.
The cosmic-ray ionization rate was assumed to be $1.3 \times 10^{-17}$ s$^{-1}$.
The ratio between diffusion energy and binding energy was set at 0.5 \citep{2006A&A...457..927G}.
The binding energies of major species and key species in the following sections are summarized in Appendix \ref{sec:bind}.
The competitive mechanism was used for surface reactions with chemical activation energy \citep{HM2008}.
Table \ref{tab:ie} lists the initial elemental abundances with respect to total hydrogen.
These elemental abundances correspond to the low-metal abundances, which are typically used for modeling the chemistry of dark clouds.
We assume that all of hydrogen is in the form of  H$_{2}$ at the initial stage.
There are 7646 gas-phase reactions and 498 gas-phase species,  mainly taken from the Kinetic Database for Astrochemistry (KIDA)\footnote{http://kida.obs.u-bordeaux1.fr}. 
We also included data taken from \citet{2015MNRAS.449L..16B} and \citet{2018ApJ...854..135S}.
\citet{2015MNRAS.449L..16B} proposed the new formation reactions forming methyl formate and dimethyl ether.
\citet{2018ApJ...854..135S} studied a new scheme for the synthesis of glycolaldehyde, acetic acid, and formic acid.
There are 5323 grain-surface reactions and 431 grain-surface species including suprathermal species \citep{2018PCCP...20.5359S}.
The surface reactions come mainly from \citet{2013ApJ...765...60G}, with additional data taken from \citet{2018ApJ...852...70B} and \citet{2018ApJ...857...89H}.
The reactions between CH$_{3}$OH and CH$_{2}^{\ast}$ \citep[the asterisk mark ($^{\ast}$) means the suprathermal species;][]{2018ApJ...852...70B} and between CH$_{3}$CO and CH$_{3}$ including suprathermal species \citep{2018ApJ...857...89H,2018PCCP...20.5359S} are included.
The self-shielding effects of H$_{2}$ \citep{1996A&A...311..690L}, CO \citep{2009A&A...503..323V}, and N$_{2}$ \citep{2013A&A...555A..14L} are included.  

In addition to the self-shielding effects, Nautilus allows for variation of different parameters through switches. 
Included switches are as follows: enhancement of H$_{2}$ grain formation (off); enabling of photodesorption of ices (on); inclusion of cosmic-ray diffusion (on)\footnote{The effect of cosmic ray impacts which cause a stochastic heating of the dust particles allowing for surface radicals to diffuse quickly and react to form more complex species \citep{2014MNRAS.440.3557R}.} ; inclusion of Eley-Rideal mechanism (off); enabling of radiolysis (off).
In our reaction network, reactions between cyanopolyynes and radicals on dust surface, which lead to destruction of cyanopolyynes, are not included. 
Since few of these processes have been studied to the best of our knowledge, and since theoretical treatments are not readily available, we did not try to add such reactions to our network. 
As shown in the following sections, cyanopolyynes in the gas-phase are destroyed mainly by atomic and molecular ions, which are not included in dust surface and ice mantles: all dust-surface and ice-mantle species are neutral.
One neutral species, which can destroy the cyanopolyynes, is neutral atomic carbon, but its abundance on dust surfaces is rather low due to rapid hydrogenation reactions even at low temperatures.

\floattable
\begin{deluxetable}{cc}
\tabletypesize{\scriptsize}
\tablecaption{Initial elemental abundances with respect to total Hydrogen \label{tab:ie}}
\tablewidth{0pt}
\tablehead{
\colhead{Element} &  \colhead{Abundance}
}
\startdata
H$_{2}$ & 0.5 \\
He & 0.09 \\
C$^{+}$ & $7.3 \times 10^{-5}$ \\
N & $2.14 \times 10^{-5}$ \\
O & $1.76 \times 10^{-4}$ \\
F & $1.8 \times 10^{-8}$ \\
Si$^{+}$ & $8 \times 10^{-9}$ \\
S$^{+}$ & $8 \times 10^{-8}$ \\
Fe$^{+}$ & $3 \times 10^{-9}$ \\
Na$^{+}$ & $2 \times 10^{-9}$ \\
Mg$^{+}$ & $7 \times 10^{-9}$ \\
Cl$^{+}$ & $1 \times 10^{-7}$ \\
P$^{+}$ & $2 \times 10^{-10}$ \\
\enddata
\tablecomments{Taken from the AL model in \citet{2017ApJ...850..105A}.}
\end{deluxetable}	

We adopted the physical evolution of the two-stage hot-core model described by \citet{2006A&A...457..927G}.
The first stage corresponds to the free-fall collapse period.
The initial gas density is $n_{\rm {H}} = 10^{4}$ cm$^{-3}$ and increases to 10$^{7}$ cm$^{-3}$ during the free-fall collapse, which lasts for $\sim 5 \times 10^{5}$ yr, while the visual extinction ($A_{V}$) increases from 5 mag to over 500 mag according to the increase in $n_{\rm {H}}$.
The temperature is kept constant at 10 K \citep{2006A&A...457..927G}.

\begin{figure}
\figurenum{1}
\plotone{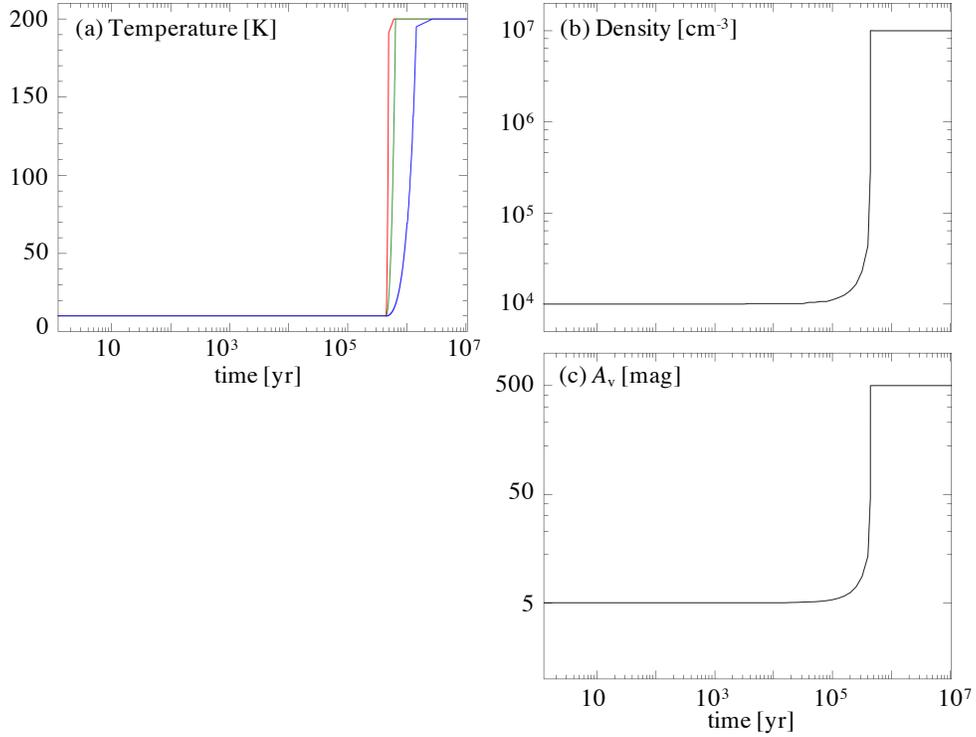}
\caption{Time variations of (a) temperature, (b) H$_{2}$ density, and (c) visual extinction, respectively. In panel (a), red, green, and blue lines indicate the heating timescales  of the Fast, Medium, and Slow warm-ups, respectively. In the other panels, the slow model was utilized. \label{fig:fa1}}
\end{figure}

The second stage is the warm-up period, during which the
 temperature rises from 10 K to 200 K according to the following formula \citep{2006A&A...457..927G}: 
\begin{equation} \label{equ:tem}
T = T_{0} + (T_{\rm {max}} - T_{0}) \Bigl( \frac{\Delta t}{t_{\rm {h}}} \Bigr)^{n},
\end{equation}
where $T_{0}$, $T_{\rm {max}}$, $t_{\rm {h}}$, and $n$ are the initial temperature (10 K), the maximum temperature (200 K), the heating timescale, and the order of the heating, respectively.
We chose $n=2$, following \citet{2008ApJ...681.1385H}.
Three heating timescales were adopted: $5 \times 10^{4}$ yr (Fast), $2 \times 10^{5}$ yr (Medium), and $1 \times 10^{6}$ yr (Slow), approximating high-mass, intermediate-mass, and low-mass star-formation, respectively \citep{2004MNRAS.354.1141V}.
The dust temperature is assumed to be equal to the gas temperature.
Figure \ref{fig:fa1} shows the time variation of temperature, H$_{2}$ density, and visual extinction.  
After the final temperature of 200 K is reached, the system remains at this temperature, at which chemistry can still occur.  
In practice, several dynamical changes such as the molecular outflow and disk formation occur, but our models do not include the effects from these phenomena.
Table \ref{tab:model} summarizes the models utilized. 
 
\floattable
\begin{deluxetable}{cccc}
\tabletypesize{\scriptsize}
\tablecaption{Summary of models \label{tab:model}}
\tablewidth{0pt}
\tablehead{
\colhead{Model} & \colhead{$t_{\rm {h}}$\tablenotemark{a}} & \colhead{phase}
}
\startdata
M1 & Fast & 3 phase \\
M2 & Medium & 3 phase \\
M3 & Slow & 3 phase \\
M4 & Fast & 2 phase \\
M5 & Medium & 2 phase \\
M6 & Slow & 2 phase \\
\enddata
\tablenotetext{a}{Fast, Medium, Slow correspond to $5 \times 10^{4}$, $2 \times 10^{5}$, and $1 \times 10^{6}$ yr, respectively.}
\end{deluxetable}	

\section{Results} \label{sec:res}

\subsection{Cyanopolyyne Chemistry during the Free-Fall Collapse Period} \label{sec:res1}

\begin{figure}
\figurenum{2}
\plotone{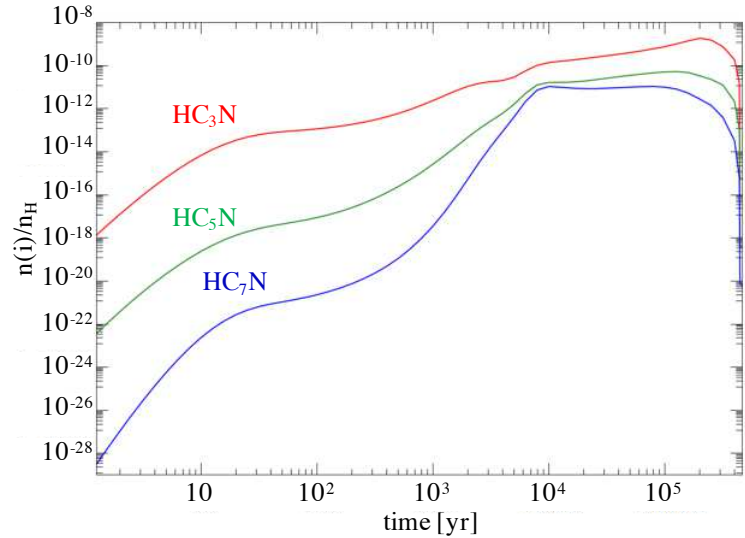}
\caption{Cyanopolyyne's abundances in the gas phase during the free-fall collapse period. The red, green, and blue lines indicate HC$_{3}$N, HC$_{5}$N, and HC$_{7}$N, respectively. \label{fig:f1}}
\end{figure}

Figure \ref{fig:f1} shows the abundances of HC$_{3}$N, HC$_{5}$N, and HC$_{7}$N in the gas phase during the free-fall collapse period.
The abundances of HC$_{3}$N and HC$_{5}$N at time scales of $5 \times 10^{3}$ -- $4 \times 10^{5}$ yr agree with the abundances observed in high-mass starless cores \citep[$X$(HC$_{3}$N)$\sim 10^{-11} - 10^{-10}$ and $X$(HC$_{5}$N)$\sim 10^{-12} - 10^{-11}$;][]{2018ApJ...854..133T}.
High-mass starless cores typically have gas densities of order 10$^{4}$ cm$^{-3}$ at fractions of a parsec \citep{2002ApJ...566..945B}. 
Therefore, the results during the free-fall collapse period seem to be reasonable.

During the free-fall collapse stage, the following reaction significantly contributes to the formation of HC$_{3}$N;
\begin{equation} \label{equ:HC3Nfree} 
{\rm {N}} + {\rm {C}}_{4}{\rm {H}} \rightarrow {\rm {HC}}_{3}{\rm {N}} + {\rm {C}}.
\end{equation}
After an interval of $6 \times 10^{3}$ yr, HC$_{3}$N is also formed significantly by the dissociative recombination reaction
\begin{equation} \label{equ:HC3Nfree3} 
{\rm {HC}}_{3}{\rm {NH}}^{+} + {\rm {e}}^{-} \rightarrow {\rm {HC}}_{3}{\rm {N}} + {\rm {H}.}
\end{equation}

Several reactions contribute to the formation of HC$_{5}$N and HC$_{7}$N, and their fraction changes with time.
The following types of reactions largely form these species:
\begin{equation} \label{equ:ion}
{\rm {H}}_{2}{\rm {C}}_{n}{\rm {N}}^{+} + {\rm {e}}^{-} \rightarrow {\rm {HC}}_{n}{\rm {N}} + {\rm {H}},
\end{equation}
\begin{equation} \label{equ:nega}
{\rm {N}} + {\rm {C}}_{n}{\rm {H}}^{-} \rightarrow {\rm {HC}}_{n}{\rm {N}} + {\rm {e}}^{-},
\end{equation}
and
\begin{equation} \label{equ:neu}
{\rm {N}} + {\rm {C}}_{n+1}{\rm {H}} \rightarrow {\rm {HC}}_{n}{\rm {N}} + {\rm {C}},
\end{equation}
where $n=5, 7$.

Cyanopolyynes are mainly destroyed by reactions with C$^{+}$ before $5 \times 10^{3}$ yr and reactions with C between $5 \times 10^{3}$ and $2 \times 10^{5}$ yr:
\begin{equation} \label{equ:des1}
{\rm {HC}}_{n}{\rm {N}} + {\rm {C}}^{+} \rightarrow {\rm {H}} + {\rm {C}}_{n+1}{\rm {N}}^{+},
\end{equation}
\begin{equation} \label{equ:des2}
{\rm {HC}}_{n}{\rm {N}} + {\rm {C}}^{+} \rightarrow {\rm {CN}} + {\rm {C}}_{n}{\rm {H}}^{+},
\end{equation}
and 
\begin{equation} \label{equ:des3}
{\rm {HC}}_{n}{\rm {N}} + {\rm {C}} \rightarrow {\rm {H}} + {\rm {C}}_{n+1}{\rm {N}},
\end{equation}
where $n=3,5,7$.

\subsection{Cyanopolyyne Chemistry during the Warm-up period} \label{sec:res2}

\begin{figure}
\figurenum{3}
 \begin{center}
  \includegraphics[bb=0 20 348 668, scale=0.7]{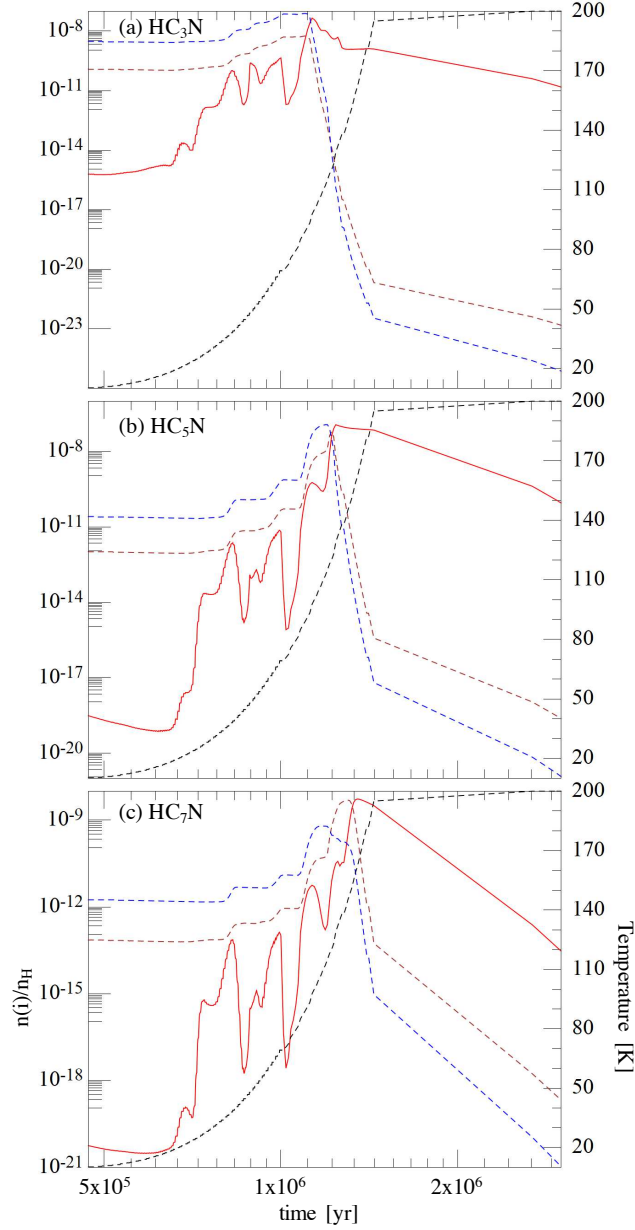}
 \end{center}
\caption{Cyanopolyyne abundances during the warm-up period of the slow three-phase model (Model M3). Panels (a), (b), and (c) show the results for HC$_{3}$N, HC$_{5}$N, and HC$_{7}$N, respectively. The red solid line, brown dashed line, and blue dashed line indicate abundances in the gas phase, dust surface, and bulk of the icy mantle, respectively. The black dashed lines indicate temperature. \label{fig:f2}}
\end{figure}

\begin{figure}
\figurenum{4}
 \begin{center}
  \includegraphics[bb=0 25 529 260, scale=0.7]{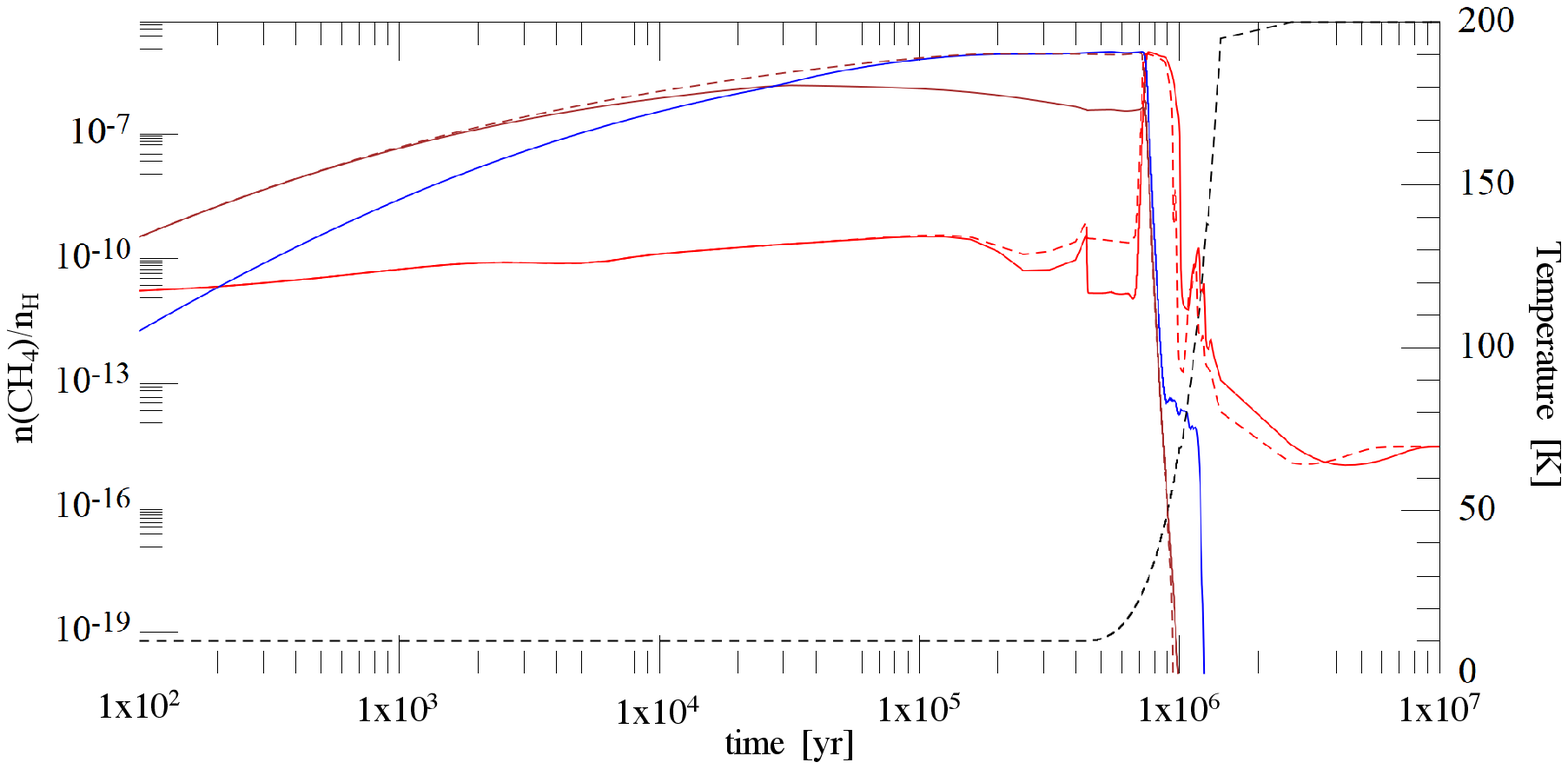}
 \end{center}
\caption{Methane (CH$_{4}$) abundances of the slow three-phase model (Model M3) and two-phase model (Model 6) during the free-fall and warm-up periods. The solid and dashed lines show the results of Model M3 and M6, respectively. The red, brown, and blue lines indicate abundances in the gas phase, dust surface, and bulk of the icy mantle, respectively. The abundance in the bulk of the icy mantle is plotted only for Model M3. The black dashed lines indicate temperature. \label{fig:fadd1}}
\end{figure}

Figure \ref{fig:f2} shows  cyanopolyyne abundances in the gas phase, dust surface, and bulk of the icy mantle during the warm-up period for the slow, three-phase model M3.
The zero of time is the starting time of the model calculation in the free-fall collapse stage.
The left-hand side of each abscissa corresponds to the start of the warm-up period, at around $5 \times 10^{5}$ yr.  
Note that the temperature is followed with time by a black dashed line.
The formation mechanisms of cyanopolyynes are basically the same in all of the models, depending on the temperature.
We use Model M3  because we can investigate the temperature dependence of cyanopolyyne chemistry during the warm-up period in most detail.

\subsubsection{HC$_{3}$N} \label{sec:resHC3N}

The HC$_{3}$N abundance, especially in the gas phase, begins an increase around $t =7 \times 10^{5}$ yr when the temperature reaches 25 K and the WCCC chemistry onsets \citep{2013ChRv..113.8981S}.
Figure \ref{fig:fadd1} shows the CH$_{4}$ abundances in the gas phase, dust surface, and bulk of the icy mantle of Models M3 and M6.
The gas-phase CH$_{4}$ abundance steeply increases and the abundances in dust surface and ice mantles decrease at $t =7 \times 10^{5}$ yr.
At that time, methane (CH$_{4}$) desorbs into the gas phase from the dust surface, and then CH$_{4}$ reacts with C$^{+}$ leading to the formation of C$_{2}$H$_{2}$ via the electron recombination reactions involving C$_{2}$H$_{3}^{+}$ and C$_{2}$H$_{4}^{+}$ \citep{2008ApJ...681.1385H}. 
The C$^{+}$ ion is formed by the reaction between CO and He$^{+}$.
C$_{2}$H$_{2}$ reacts with CN to form HC$_{3}$N:
\begin{equation} \label{equ:r1}
{\rm {C}}_{2}{\rm {H}}_{2} + {\rm {CN}} \rightarrow {\rm {HC}}_{3}{\rm {N}} + {\rm {H}}.
\end{equation}
This reaction was suggested as the main formation pathway of HC$_{3}$N based on observations of its three $^{13}$C isotopologues toward the G28.28$-$0.36 MYSO \citep{2016ApJ...830..106T}.
Therefore, our model result and the observational result reach the same conclusion. 
Small subsequent peaks in the HC$_{3}$N gaseous abundance  correspond to an enhancement in Reaction (\ref{equ:r1}), which relates to both the CN and C$_{2}$H$_{2}$ abundances.
For example, the CN abundance sharply increases due to its direct desorption from dust grains around $t = 7.7 \times 10^{5}$ yr ($T \simeq 31$ K).
Around $t = 9.3 \times 10^{5}$ yr ($T \simeq 55$ K), the gas-phase C$_{2}$H$_{2}$ abundance increases  because of its desorption from dust grains.  The increases in the gas phase values of CN and C$_2$H$_2$ lead to separate enhancements in the HC$_3$N gas phase abundance through Reaction (\ref{equ:r1}).   

At these low temperatures, HC$_{3}$N formed in the gas phase is partially depleted onto dust grains following each enhancement and accumulates in the bulk of the icy mantle.
The result is a pattern for the HC$_3$N abundance vs time that resembles a spectrum.
At $t \simeq 1.1 \times 10^{6}$ yr ($T \simeq 90$ K), HC$_{3}$N sublimates from dust grains and reaches its peak abundance in the gas phase. 
The peak abundance in the gas phase is consistent with that of the bulk of the icy mantle, as shown in panel (a) of Figure \ref{fig:f2}.

\subsubsection{HC$_{5}$N} \label{sec:resHC5N}

HC$_{5}$N is mainly formed by the following reaction in the warming gas ($T > 25$ K):
\begin{equation} \label{equ:r2}
{\rm {C}}_{4}{\rm {H}}_{2} + {\rm {CN}} \rightarrow {\rm {HC}}_{5}{\rm {N}} + {\rm {H}},
\end{equation}
while the precursor C$_{4}$H$_{2}$ is produced by the reaction
\begin{equation} \label{equ:r3}
{\rm {CCH}} + {\rm {C}}_{2}{\rm {H}}_{2} \rightarrow {\rm {C}}_{4}{\rm {H}}_{2} + {\rm {H}},
\end{equation}
and the radical CCH is formed by the electron recombination reaction of C$_{2}$H$_{3}^{+}$.
The ion C$_{2}$H$_{3}^{+}$ is formed by the reaction between CH$_{4}$ and C$^{+}$, as discussed  in the preceding section.
The gas-phase species C$_{4}$H$_{2}$ sharply increases at $t = 1.0 \times 10^{6}$ yr ($T \simeq 73$ K) via its direct sublimation from dust grains.
Owing to increases in the C$_{4}$H$_{2}$ abundance, Reaction (\ref{equ:r2}) is enhanced, leading to an increase in the HC$_{5}$N  abundance. 
The radical CCH also reacts with HC$_{3}$N in the gas phase to produce HC$_{5}$N by the reaction
\begin{equation} \label{equ:r4}
{\rm {CCH}} + {\rm {HC}}_{3}{\rm {N}} \rightarrow {\rm {HC}}_{5}{\rm {N}} + {\rm {H}}.
\end{equation}

The species HC$_{5}$N formed in the gas phase is partially accreted into the bulk of the icy mantle, as is the case for HC$_{3}$N.
At $t = 1.2 \times 10^{6}$ yr ($T \simeq 115$ K), HC$_{5}$N is desorbed into the gas, and reaches its peak abundance.
This abundance corresponds to that in the bulk of the mantle, right before desorption, as shown in panel (b) of Figure \ref{fig:f2}, in a similar manner to  HC$_{3}$N.
The subsequent decrease is caused by  reaction with HCO$^{+}$ and other protonated ions.

\subsubsection{HC$_{7}$N} \label{sec:resHC7N}

After the temperature reaches 25 K, HC$_{7}$N is efficiently formed by the following reaction:
\begin{equation} \label{equ:r5}
{\rm {C}}_{6}{\rm {H}}_{2} + {\rm {CN}} \rightarrow {\rm {HC}}_{7}{\rm {N}} + {\rm {H}}.
\end{equation}
The C$_{6}$H$_{2}$ species is formed by the reaction between CCH and C$_{4}$H$_{2}$, which is similar to Reaction (\ref{equ:r3}).
In addition, gaseous C$_{6}$H$_{2}$ reaches its peak abundance just after its direct sublimation from dust grains at $t=1.2 \times 10^{6}$ yr ($T \simeq 94$ K) and the reaction rate of Reaction (\ref{equ:r5}) increases.
This leads to an increase in the HC$_{7}$N abundance. 
The  HC$_{7}$N species is also produced by the reaction
\begin{equation} \label{equ:r6}
{\rm {CCH}} + {\rm {HC}}_{5}{\rm {N}} \rightarrow {\rm {HC}}_{7}{\rm {N}} + {\rm {H}}.
\end{equation}

Analogously to the cases of HC$_{3}$N and HC$_{5}$N, HC$_{7}$N  also accumulates on the dust surface and in the bulk of the icy mantle and is desorbed into the gas phase around $t = 1.3 \times 10^{6}$ yr ($T \simeq 145$ K).
The peak abundance of HC$_{7}$N in the gas phase is comparable to that on the dust surface, not in the bulk of the icy mantle, as shown in panel (c) of Figure \ref{fig:f2}.

Besides adsorption onto dust grains, cyanopolyynes are efficiently destroyed by the reactions with HCO$^{+}$ and other ions during the warm-up period:
\begin{equation} \label{equ:rhco}
{\rm {HC}}_{n}{\rm {N}} + {\rm {HCO}}^{+} \rightarrow {\rm {H}}_{2}{\rm {C}}_{n}{\rm {N}}^{+} + {\rm {CO}},
\end{equation}
where $n = 3,5,7$.

The peak abundance of HC$_{5}$N is higher than that of HC$_{3}$N, which may not be realistic.
Occasionally the largest species in a sequence, such as HC$_{7}$N (if it is truly the largest) is not described well in a network.
The high HC$_{5}$N abundance may reflect that there are unknown important destruction reactions of cyanopolyynes both in the gas phase and dust surface.
For example, radicals can move on dust grains during the warm-up period and reactions between cyanopolyynes and radicals may decrease the abundances of cyanopolyynes on dust surface and ice mantles.
However, our current reaction networks does not contain such reactions, while photodissociation destruction processes are included.
Besides photodissociation processes, if other destruction reactions of cyanopolyynes on dust surface exist, the longer cyanopolyynes will be more affected because longer cyanopolyynes stay longer time due to higher sublimation temperatures in our current model.

In summary, the cyanopolyynes are produced by a combination of neutral-neutral and ion-neutral gas-phase  reactions during the warm-up period ($T > 25$ K) and accumulate on and in the dust mantles before the temperature reaches their sublimation temperatures.
The sublimation of first CN and secondly the C$_{2n}$H$_{2}$ ($n=1,2,3$) species enhances key reactions  to form the cyanopolyynes, which partly accrete onto dust mantles.
As the temperature rises, the pattern of enhancements in the production of the gaseous  cyanopolyynes followed by  partial accretion onto grains leads to a characteristic spectral-type pattern.
When the temperature rises past the sublimation temperatures of cyanopolyynes, they desorb into the gas, and reach their peak abundances.  
The spectral-type abundance profile is best seen in the slow models, but exists to a lesser extent in the faster models too.
Cyanopolyynes are mainly destroyed by reactions with HCO$^{+}$ and other protonated ions in the gas phase and their abundances decrease.

\section{Discussion} \label{sec:dis}

\subsection{Comparisons of Cyanopolyyne Abundances among Models} \label{sec:dis1}

\begin{figure}
\figurenum{5}
 \begin{center}
  \includegraphics[bb=0 20 554 574, scale=0.85]{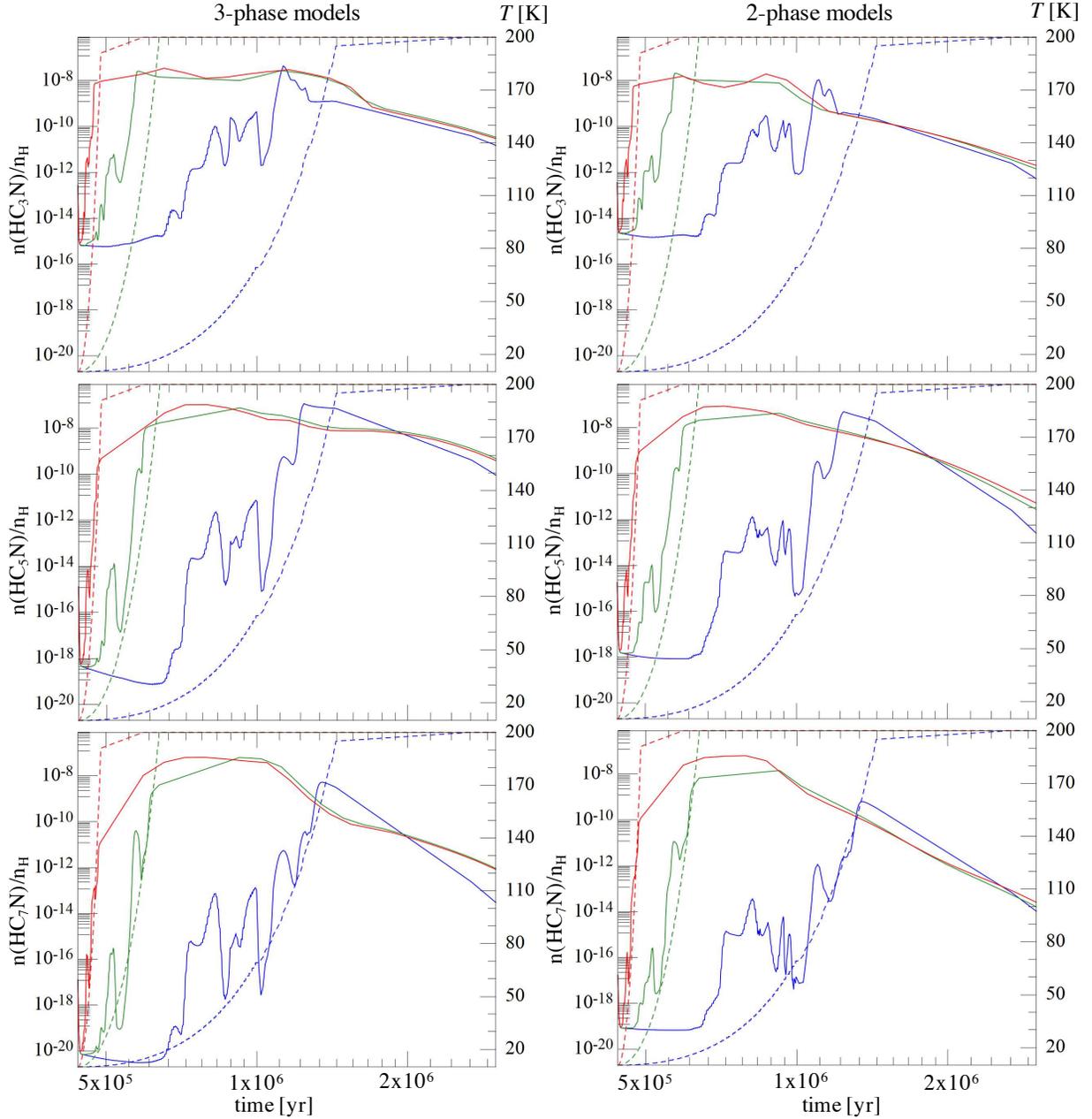}
 \end{center}
\caption{Comparisons of gas-phase cyanopolyyne abundances in the warm-up period among models. The upper, middle, and lower panels show results of HC$_{3}$N, HC$_{5}$N, and HC$_{7}$N, respectively. The different colors of lines indicate different heating timescales; red, green, and blue indicate Fast, Medium, and Slow, respectively. The dashed lines indicate the temperature and their colors correspond to the heating timescales. The left panels show the 3-phase models (M1 -- M3) and right panels show the 2-phase models (M4 -- M6). \label{fig:f3}}
\end{figure}

\floattable
\begin{deluxetable}{ccccccccc}
\tabletypesize{\scriptsize}
\tablecaption{The peak calculated abundances of cyanopolyynes in the gas phase and the corresponding temperature \label{tab:cyano_peak}}
\tablewidth{0pt}
\tablehead{
\colhead{Model} & \multicolumn{2}{c}{HC$_{3}$N} & \colhead{} & \multicolumn{2}{c}{HC$_{5}$N} & \colhead{} & \multicolumn{2}{c}{HC$_{7}$N} \\
\cline{2-3}\cline{5-6}\cline{8-9}
\colhead{} & \colhead{Abundance} & \colhead{$T$ (K)} &  \colhead{} & \colhead{Abundance} & \colhead{$T$ (K)} & \colhead{} & \colhead{Abundance} & \colhead{$T$ (K)}
}
\startdata		
M1 & $3.62 \times 10^{-8}$ & 200 & & $1.08 \times 10^{-7}$ & 200 & & $6.03 \times 10^{-8}$ & 200 \\
M2 & $2.79 \times 10^{-8}$ &200 & & $7.66 \times 10^{-8}$ & 200 & & $6.20 \times 10^{-8}$ & 200 \\
M3 & $4.39 \times 10^{-8}$ & 98 & & $1.14 \times 10^{-7}$ & 129 & & $5.19 \times 10^{-9}$ & 162 \\
M4 & $1.98 \times 10^{-8}$ & 200 & & $8.97 \times 10^{-8}$ & 200 & & $6.41 \times 10^{-8}$ & 200 \\
M5 & $2.15 \times 10^{-8}$ & 96 & & $4.42 \times 10^{-8}$ & 200 & & $1.45 \times 10^{-8}$ & 200 \\
M6 & $1.11 \times 10^{-8}$ & 92 & & $5.02 \times 10^{-8}$ & 129 & & $6.44 \times 10^{-10}$ & 162 \\
\enddata
\end{deluxetable}	

Figure \ref{fig:f3} shows comparisons of the cyanopolyyne abundances in the gas-phase during the warm-up period, among all models, while Table \ref{tab:cyano_peak} summarizes the peak abundances during the warm-up and hot-core periods of cyanopolyynes in the gas phase and the corresponding temperatures in each model.  
In the fast and medium warm-up  models (red and green lines), cyanopolyynes reach higher abundances in the gas phase after their desorption.
Given the rapidity of desorption,  their peak abundances do not correspond to their abundances on the dust surface or bulk of the icy mantle during the desorption.  
The post-desorption syntheses in the gas phase are particularly important for the longer chains.
One possible reason for their importance  is that there is not enough time to accumulate on dust surfaces or in the bulk of the icy mantles during the short heating timescale.  
On the other hand, accumulation on dust surface and in the bulk of the icy mantle play essential roles in producing the high cyanopolyyne abundances during the warm-up period when there is enough time for cyanopolyynes to adsorb onto dust grains.
This seems to depend on the assumed initial density.

When we compare the peak abundances in Model M3 (slow; 3-phase) with those in Model M6 (slow, 2-phase), we find that the peak abundances in M3 are higher  by a factor of 2 -- 8, as shown in Table \ref{tab:cyano_peak}.
These different peak abundances between Models M3 and M6 support the hypothesis that accumulation of cyanopolyynes in the bulk of the icy mantle enhances their abundances in the gas phase after their desorption at higher temperatures, because, basically, the ice prevents the gas phase destruction of cyanopolyynes by maintaining their peak abundances unaltered in the mantles.

\subsection{Comparisons with Observations} \label{sec:dis2}

Table \ref{tab:obs} contains summaries of the observed abundances of HC$_{5}$N, CH$_{3}$CCH, and CH$_{3}$OH in three MYSOs \citep{2017ApJ...844...68T,2018ApJ...866..150T} and the peak calculated abundances in each model.
We excluded HC$_{3}$N because its column densities and excitation temperatures in the observed MYSOs seem to include large uncertainties \citep{2018ApJ...866..150T}.
These observations were carried out using  single-dish telescopes and thus, because of beam dilution, the derived abundances and rotational temperatures are  lower limits \citep{2017ApJ...844...68T}. 
Hence, our criterion for reproduction is that the model produces more than the observed lower limit in the following sections.

We found that the maximum model abundances of the three gas-phase species are higher than the observed lower limits in all of the models.
We also derived their abundances in L1527, a low-mass WCCC source,  and found, once again that the model peak abundances exceed the observed lower limits.  
\floattable
\begin{deluxetable}{cccc}
\tabletypesize{\scriptsize}
\tablecaption{Lower limits of observed abundances in MYSOs and the peak calculated abundances  of HC$_{5}$N, CH$_{3}$CCH, and CH$_{3}$OH \label{tab:obs}}
\tablewidth{0pt}
\tablehead{
\colhead{Source/Model} & \colhead{HC$_{5}$N} & \colhead{CH$_{3}$CCH} & \colhead{CH$_{3}$OH}
}
\startdata
{\it {Observation}}\tablenotemark{a} & & & \\
G12.89+0.49 & $5.0^{+3.1}_{-2.7} \times 10^{-10}$ & $2.1^{+1.4}_{-1.1} \times 10^{-8}$ & $6.0^{+9.0}_{-4.5} \times 10^{-8}$ \\
G16.86$-$2.16 & ($8.0 \pm 3.5) \times 10^{-10}$ & $1.6^{+0.95}_{-0.75} \times 10^{-8}$ & $4.4^{+5.5}_{-2.9} \times 10^{-8}$ \\
G28.28$-$0.36 & $2.1^{+1.3}_{-1.0} \times 10^{-9}$ & $3.8^{+2.6}_{-2.2} \times 10^{-8}$ & $2.3^{+3.5}_{-1.7} \times 10^{-8}$ \\
L1527 (WCCC)\tablenotemark{b} & ($1.2 \pm 0.3$)$\times 10^{-10}$ & ($1.07 \pm 0.05$)$\times 10^{-9}$ & ($1.1 \pm 0.2$)$\times 10^{-9}$ \\
& & &  \\
{\it {Model}} & & & \\
M1 & $1.08 \times 10^{-7}$ & $9.27 \times 10^{-7}$ & $6.32 \times 10^{-6}$ \\
M2 & $7.66 \times 10^{-8}$ & $9.23 \times 10^{-7}$ & $4.76 \times 10^{-6}$ \\
M3 & $1.14 \times 10^{-7}$ & $2.42 \times 10^{-7}$ & $9.66 \times 10^{-8}$ \\
M4 & $8.97 \times 10^{-8}$ & $9.61 \times 10^{-7}$ & $3.60 \times 10^{-6}$ \\
M5 & $4.42 \times 10^{-8}$ & $9.03 \times 10^{-7}$ & $3.02 \times 10^{-6}$ \\
M6 & $5.02 \times 10^{-8}$ & $2.80 \times 10^{-7}$ & $6.46 \times 10^{-8}$ \\
\enddata
\tablenotetext{a}{Abundances taken from \citet{2017ApJ...844...68T,2018ApJ...866..150T} and converted to abundances with respect to total hydrogen.}
\tablenotetext{b}{Column densities of HC$_{5}$N, CH$_{3}$CCH, and CH$_{3}$OH taken from \citet{2013ChRv..113.8981S} and $N$(H$_{2}$) value of $2.8 \times 10^{22}$ cm$^{-2}$ taken from \citet{2002AA...389..908J}. The abundances with respect to H$_{2}$ are converted to abundances with respect to total hydrogen.}
\end{deluxetable}	

\begin{figure}
\figurenum{6}
 \begin{center}
  \includegraphics[bb=0 20 547 573, scale=0.85]{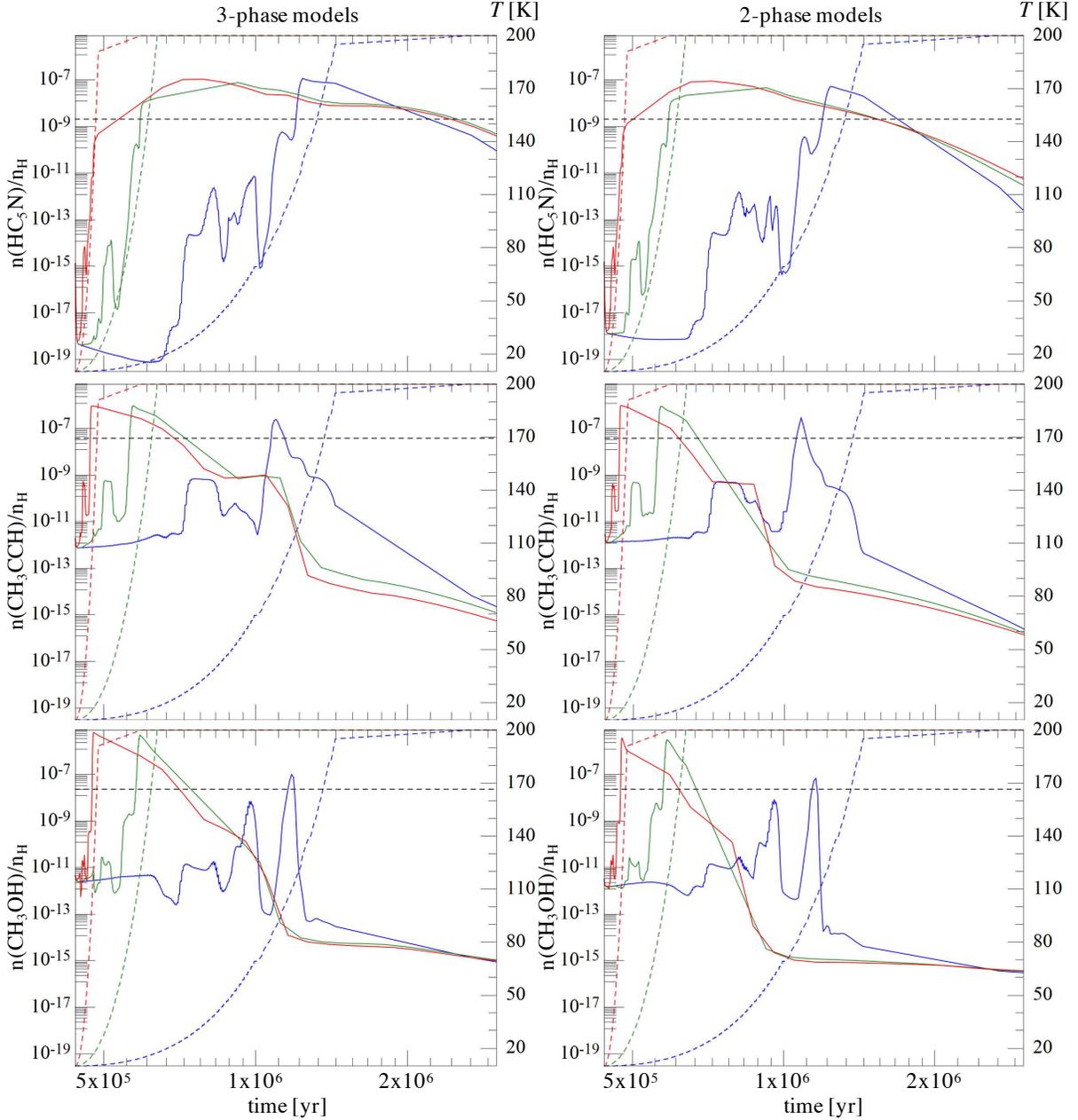}
 \end{center}
\caption{Comparisons between model results and lower limits derived for G28.28$-$0.36 (horizontal black dashed lines). The upper, middle, and lower panels show results for gas-phase HC$_{5}$N, CH$_{3}$CCH, and CH$_{3}$OH, respectively. The different colors of lines indicate different heating timescales; red, green, and blue indicate Fast, Medium, and Slow, respectively. The dashed curves indicate the temperature and their colors correspond to the heating timescales. \label{fig:f4}}
\end{figure}

\floattable
\begin{deluxetable}{cccccccc}
\tabletypesize{\scriptsize}
\tablecaption{The range of age and lower limit of temperature constrained by comparisons with the observational results in G28.28$-$0.36 \label{tab:G28}}
\tablewidth{0pt}
\tablehead{
\colhead{} & \multicolumn{3}{c}{Age (yr)} & \colhead{} & \multicolumn{3}{c}{Lower limit of temperature (K)} \\
\cline{2-4} \cline{6-8}
\colhead{Model} & \colhead{HC$_{5}$N} & \colhead{CH$_{3}$CCH} & \colhead{CH$_{3}$OH} & \colhead{} & \colhead{HC$_{5}$N} & \colhead{CH$_{3}$CCH} & \colhead{CH$_{3}$OH}
}
\startdata
M1 & $5.94 \times 10^{5} - 2.48 \times 10^{6}$ & $4.71 \times 10^{5} - 7.19 \times 10^{5}$ & $4.75 \times 10^{5} - 7.19 \times 10^{5}$ & & 200 & 84 & 100 \\
M2 & $5.92 \times 10^{5} - 2.64 \times 10^{6}$ & $5.64 \times 10^{5} - 6.39 \times 10^{5}$ & $5.79 \times 10^{5} - 6.39 \times 10^{5}$ & & 120 & 84 & 100 \\ 
M3 & $1.21 \times 10^{6} - 1.44 \times 10^{6}$ & $1.07 \times 10^{6} - 1.16 \times 10^{6}$ & $1.16 \times 10^{6} - 1.19 \times 10^{6}$ & & 118 & 82 & 105 \\
M4 & $5.94 \times 10^{5} - 1.70 \times 10^{6}$ & $4.70 \times 10^{5} - 5.94 \times 10^{5}$ & $4.74 \times 10^{5} - 5.94 \times 10^{5}$ & & 200 & 78 & 98 \\ 
M5 & $5.90 \times 10^{5} - 1.64 \times 10^{6}$ & $5.59 \times 10^{5} - 6.39 \times 10^{5}$ & $5.74 \times 10^{5} - 6.39 \times 10^{5}$ & & 118 & 78 & 96 \\ 
M6 & $1.19 \times 10^{6} - 1.44 \times 10^{6}$ & $1.05 \times 10^{6} - 1.12 \times 10^{6}$ & $1.13 \times 10^{6} - 1.17 \times 10^{6}$ & & 112 & 78 & 98 \\
\enddata
\end{deluxetable}	

We are particularly interested in G28.28$-$0.36, because this source shows a uniquely  high HC$_{5}$N/CH$_{3}$OH feature, discussed below, among the observed sources \citep{2018ApJ...866..150T}.  In addition, G28.28$-$0.36 shows a relatively high HC$_{5}$N abundance, as derived with the  single-dish observations \citep{2017ApJ...844...68T}.
In the Orion KL hot core, the HC$_{5}$N abundance with respect to H$_{2}$ was derived to be $1.7 \times 10^{-10}$ \citep{2013A&A...559A..51E}.
Therefore, the HC$_{5}$N abundance in G28.28$-$0.36 is higher than that in the Orion KL hot core by more than one order of magnitude.
As  already mentioned in Section \ref{sec:intro}, the HC$_{5}$N abundance in G28.28$-$0.36 is higher than that in L1527 by a factor of 20, as shown in Table \ref{tab:obs}.
The particularly high HC$_{5}$N abundance in G28.28$-$0.36 cannot be explained by warm carbon-chain chemistry (WCCC) in the lukewarm envelope \citep[$T \sim 20-30$ K,][]{2008ApJ...681.1385H,2011ApJ...743..182H}.

We now compare our model results with the lower limits of HC$_{5}$N, CH$_{3}$CCH, and CH$_{3}$OH in G28.28$-$0.36 in order to constrain the minimum temperatures in the warm-up period where they reside.
Figure \ref{fig:f4} shows the model results with the lower limits to molecular abundances in G28.28$-$0.36 as horizontal black dashed lines.
Table \ref{tab:G28} summarizes the range of ages and lower limits of temperature constrained by the comparisons between the model results and the observed lower limits. 
The range of ages is determined as the continuous times when the model abundances are higher than the observed lower limits.

The observed lower limit of HC$_{5}$N is reproduced only at ages after it desorbs from dust grains in all the models.
Therefore, the chemistry in the lukewarm temperature  range of $25 < T < 100$ K cannot explain the high HC$_{5}$N abundance in G28.28$-$0.36, because its desorption temperature is higher than 100 K. 
The ages when the temperature reaches at the HC$_{5}$N sublimation temperature of $\sim 115$ K are $5.9 \times 10^{5}$ yr and $1.2 \times 10^{6}$ yr in the Medium and Slow warm-up models, respectively.
In the Fast warm-up models, further gas-phase formation is needed to reproduce the observed HC$_{5}$N abundance.
Hence, the lower limit of the observed HC$_{5}$N abundance cannot be reproduced at its sublimation temperature ($t \sim 4.8 \times 10^{5}$ yr).

The carbon-chain species CH$_{3}$CCH,  shown in the middle panels of Figure \ref{fig:f4}, is formed in the gas phase during the warm-up period triggered initially by desorption of CH$_{4}$ via the following reactions:
\begin{equation} \label{rea:CH3CCH1}
{\rm {CH}}_{4} + {\rm {C}}^{+} \rightarrow {\rm {C}}_{2}{\rm {H}}_{3}^{+} + {\rm {H}},
\end{equation}
\begin{equation} \label{rea:CH3CCH2}
{\rm {CH}}_{4} + {\rm {C}}_{2}{\rm {H}}_{3}^{+} \rightarrow {\rm {C}}_{3}{\rm {H}}_{5}^{+} + {\rm {H}}_{2},
\end{equation}
followed by
\begin{equation} \label{rea:CH3CCH3}
{\rm {C}}_{3}{\rm {H}}_{5}^{+} + {\rm {e}}^{-} \rightarrow {\rm {CH}}_{3}{\rm {CCH}} + {\rm {H}}.
\end{equation}
The CH$_{3}$CCH is subsequently adsorbed onto dust grains, in analogy with the cyanopolyynes, and starts to desorb at temperature of 70 K. 
The minimum temperature at which its lower limit observed in G28.28$-$0.36 ($3.8^{+2.6}_{-2.2} \times 10^{-8}$) is reproduced occurs only after its desorption from dust grains at $\approx$ 80 K, corresponding to ages of $4.7 \times 10^{5}$ yr, $5.6 \times 10^{5}$ yr, and $1.1 \times 10^{6}$ yr in the Fast, Medium, and Slow warm-up models, respectively.
Its abundances decrease more rapidly compared to those of HC$_{5}$N, and the ranges of time which can reproduce the observed abundance become narrower.
Unlike HC$_{5}$N, CH$_{3}$CCH is destroyed by the reaction with atomic oxygen, leading to CO + C$_{2}$H$_{4}$.
This reaction seems to contribute to the faster destruction of CH$_{3}$CCH.
 
The lower panels of Figure \ref{fig:f4} show the results for CH$_{3}$OH, which is not a species produced by WCCC, because its sole production occurs on the grains via hydrogenation of CO.
Nevertheless, its lower limit observed in G28.28$-$0.36 is also reproduced only after it desorbs into the gas phase at temperatures above $\sim 100$ K.
The corresponding ages of the CH$_{3}$OH sublimation are $4.8 \times 10^{5}$ yr, $5.8 \times 10^{5}$ yr, and $1.1 \times 10^{6}$ yr in the Fast, Medium, and Slow warm-up models, respectively.
Methanol is destroyed by reactions with ions such as HCO$^{+}$.

In summary, the lower limits of all of these species derived in G28.28$-$0.36 can be reproduced only after desorption from dust grains. 
The observed species in G28.28--0.36 seem to reside in higher temperature regions when compared with WCCC sources.
The HC$_{5}$N and CH$_{3}$OH abundances do not become higher than the observed abundances at the same age in the Slow warm-up models, although their peaks can be higher than the observed ones at different ages (Table \ref{tab:G28}).

\begin{figure}
\figurenum{7}
 \begin{center}
  \includegraphics[bb=0 25 540 283, scale=0.7]{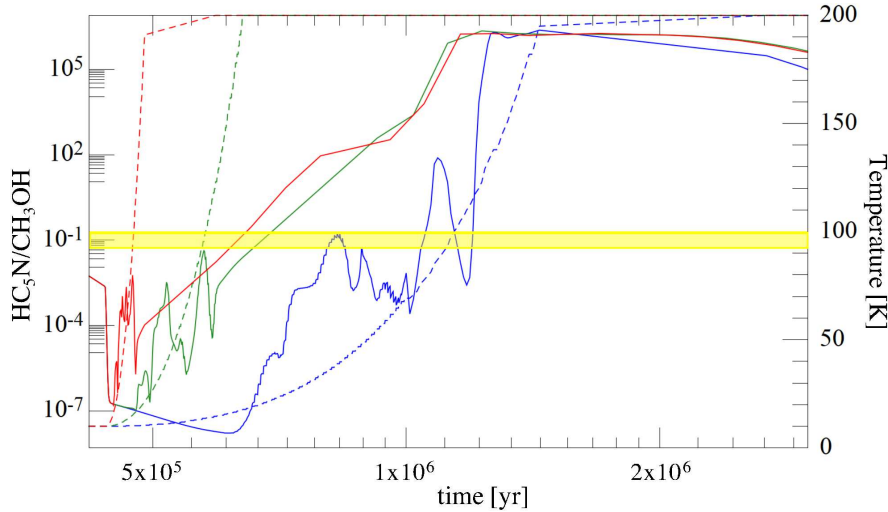}
 \end{center}
\caption{Temporal dependence of the HC$_{5}$N/CH$_{3}$OH abundance ratio after the warm-up period. The different colors of lines indicate different heating timescales; red, green, and blue indicate Fast, Medium, and Slow, respectively. The dashed curves indicate the temperature and their colors correspond to the heating timescales. The yellow range corresponds to the value observed in the G28.28-0.36 ($0.091^{+0.109}_{-0.039}$). \label{fig:fadd2}}
\end{figure}

We now return to a comparison of  theory and observation concerning the abundance ratio of HC$_5$N to methanol in G28.28--0.36.
The derived $N$(HC$_{5}$N)/$N$(CH$_{3}$OH) column density ratio in G28.28--0.36 ($0.091^{+0.109}_{-0.039}$) is higher than that in the other three MYSOs by more than one order of magnitude \citep{2018ApJ...866..150T}.
This ratio will be much lower in some HC$_{5}$N-undetected MYSOs than the HC$_{5}$N-detected MYSOs.
In fact, \citet{2014MNRAS.443.2252G} did not detect HC$_{5}$N in half of their target MYSOs. 

Figure \ref{fig:fadd2} shows temporal dependence of the HC$_{5}$N/CH$_{3}$OH abundance ratio of Models M1 -- M3.
Because the individual abundances of  both HC$_{5}$N and CH$_{3}$OH can be reproduced only after desorption from the gas, we investigated the ratio of their abundances only in this region.
The actual investigated ranges of age are $4.75 \times 10^{5}$ -- $2.48 \times 10^{6}$ yr, $5.79 \times 10^{5}$ -- $1.02 \times 10^{6}$ yr, and $1.16 \times 10^{6}$ -- $1.3 \times 10^{6}$ yr for Models M1, M2, and M3, respectively.
The closest calculated values to the observed ratios  are $\sim 0.20$ at $6 \times 10^{5}$ yr ($T = 200$ K) in Model M1 and $\sim 0.16$ at $1.2 \times 10^{6}$ yr ($T \simeq 115$ K) in Model M3.
In the case of Model M2, the calculated ratio with the current time resolution changes too rapidly and we could not find the exact value and age when the observed ratio is reproduced.
From Figure \ref{fig:fadd2}, the observed ratio may be reproduced around $8 \times 10^{5}$ yr ($T = 200$ K) in Model M2.
All of these ages correspond closely to the time when HC$_{5}$N reaches the peak abundances.
These results may suggest that desorption of HC$_{5}$N or its further gas-phase formation have just started in hot regions in G28.28--0.36.

\subsection{Effects of Cosmic Ray Ionization Rates on HC$_{5}$N Abundance} \label{sec:dis5}

\begin{figure}
\figurenum{8}
 \begin{center}
  \includegraphics[bb=0 30 479 557, scale=0.65]{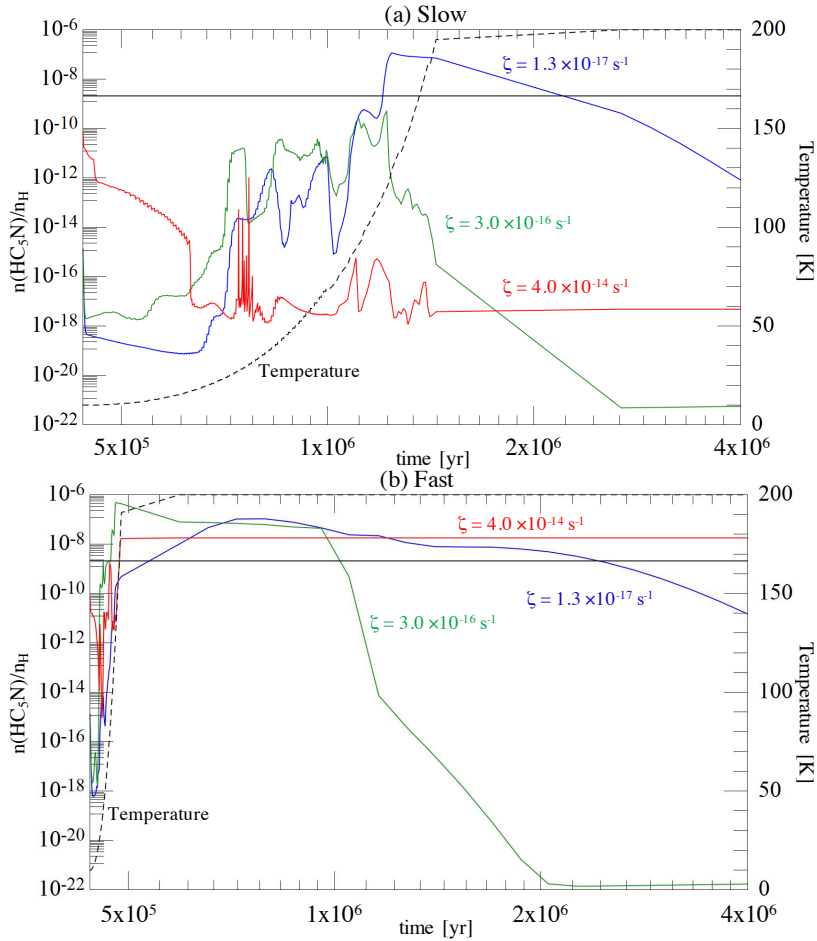}
 \end{center}
\caption{HC$_{5}$N abundances calculated with 3-phase models during the warm-up period. Panels (a) and (b) show the results using the Slow (M3) and Fast  (M1) heating timescales, respectively. The different colors of lines indicate different cosmic-ray ionization rates: blue, green, and red indicate $1.3 \times 10^{-17}$, $3.0 \times 10^{-16}$, and $4.0 \times 10^{-14}$ s$^{-1}$, respectively. The black dashed curves indicate the temperature. The horizontal black lines indicate the lower limit of the observed abundance in G28.28--0.36. \label{fig:f6}}
\end{figure}

\citet{2017A&A...605A..57F} suggested that a high cosmic-ray ionization rate of $\zeta \simeq 4 \times 10^{-14}$ s$^{-1}$ can reproduce their observational results for cyanopolyynes in OMC-2 FIR4.
The observational results of the $c$-C$_{3}$H$_{2}$ abundance were also reproduced by such a high cosmic-ray ionization rate \citep{2018ApJ...859..136F}.
In addition to runs with the common cosmic-ray ionization rate of $\zeta = 1.3 \times 10^{-17}$ s$^{-1}$, we ran our models with high cosmic-ray ionization rates of $\zeta = 3.0 \times 10^{-16}$ s$^{-1}$ and $4.0 \times 10^{-14}$ s$^{-1}$, as \citet{2017A&A...605A..57F} had done.
Such high cosmic-ray ionization rates could be possible in protostellar systems \citep{2016A&A...590A...8P}.
Our results for gas-phase HC$_{5}$N abundances using these three ionization rates are shown in Figure \ref {fig:f6}. 
We used  Models M3 and M1 with Slow and Fast warm-up timescales in panels (a) and (b), respectively.

In panel (a), the peak abundances decrease with an increase in the cosmic-ray ionization rate. 
This is caused by the fact that HC$_{5}$N is destroyed in the gas phase by reaction with H$^{+}$ and other ions before it is adsorbed onto dust grains.
The H$^{+}$ ion destroys HC$_{5}$N most efficiently after $7 \times 10^{5}$ yr:
\begin{equation} \label{rea:dest1}
{\rm {HC}}_{5}{\rm {N}}+ {\rm {H}}^{+} \rightarrow {\rm {H}} + {\rm {HC}}_{5}{\rm {N}}^{+}.
\end{equation}
This reaction is essential especially in the model with the cosmic-ray ionization rate of $4 \times 10^{-14}$ s$^{-1}$.
The HC$_{5}$N$^{+}$ ion partly reacts with H$_{2}$ to form H$_{2}$C$_{5}$N$^{+}$, which contributes to re-formation of HC$_{5}$N through the electron dissociative recombination reaction.
However, the reaction between HC$_{5}$N$^{+}$ and H$_{2}$ is much less effective compared to the electron dissociative recombination of HC$_{5}$N$^{+}$ to form C$_{5}$N and H, because the electron abundance becomes high in conditions with high cosmic-ray ionization rates.
Polyatomic ions such as HCO$^{+}$ and H$_{3}^{+}$ react with electrons, leading to dissociative recombination, more rapidly than H$^{+}$, and hence the relative abundance of H$^{+}$ to polyatomic ions becomes high. 
Thus, the H$^{+}$ becomes a reaction partner of HC$_{5}$N.
In the model with the cosmic-ray ionization rate of $3.0 \times 10^{-16}$ s$^{-1}$, on the other hand, the HCO$^{+}$ ion is the main contributor to the destruction of HC$_{5}$N after $10^{6}$ yr:
\begin{equation} \label{rea:dest2}
{\rm {HC}}_{5}{\rm {N}} + {\rm {HCO}}^{+} \rightarrow {\rm {CO}} + {\rm {H}}_{2}{\rm {C}}_{5}{\rm {N}}^{+}.
\end{equation}
The formed H$_{2}$C$_{5}$N$^{+}$ will react with electrons and partly ($\sim 45$ \%) will go back to HC$_{5}$N. 
As discussed in the previous sections, the accumulation of HC$_{5}$N onto dust surface and in ice mantles is an important process in the slow warm-up models. 
However, the HC$_{5}$N destruction by ions works efficiently in the high cosmic-ray ionization models, and the accumulation of HC$_{5}$N is suppressed, leading to the lower HC$_{5}$N abundances.
The HC$_{5}$N abundances in models with  the two higher cosmic-ray ionization rates cannot reproduce its observed abundance in G28.28--0.36 even at peak abundance.
However, there is no significant difference in peak abundance among the fast warm-up models with different cosmic-ray ionization rates in panel (b).
With the highest cosmic-ray ionization rate, HC$_{5}$N continues to be formed by Reaction (\ref{equ:r2}) efficiently, while the reaction is less efficient due to the significantly lower abundance of C$_{4}$H$_{2}$ in the models with the cosmic-ray ionization rate of $3.0 \times 10^{-16}$ s$^{-1}$ after $\sim 1 \times 10^{6}$ yr.  
This highest cosmic-ray ionization model can reproduce the observed abundance in G28.28--0.36 for all times after $4.86 \times 10^{5}$ yr ($T \geq 170$ K).

The HC$_{5}$N abundance in the slow warm-up model with cosmic-ray ionization rate of $4.0 \times 10^{-14}$ s$^{-1}$ is significantly lower than in the same cosmic-ray ionization rate model with fast warm-up. 
The difference between slow and fast warm-up models is caused by the different abundances of H$^{+}$ and H$_{3}^{+}$, which in turn are related to the abundances of H and H$_{2}$. 
In the slow warm-up model, the H$_{2}$ formation rate is slowed down by an exponential factor of the form exp($-\frac{E_{\rm {diff}}}{T}$), where $E_{\rm {diff}}$ is known as the diffusion barrier and $T$ is the temperature.
The diffusion barrier slows the average motion of the reacting hydrogen atoms.
The destruction rate of H$_{2}$ is larger than its formation rate in such a high cosmic-ray ionization condition. 
The retarding effect of the diffusion barrier is lessened in the fast warm-up model because of the higher temperatures even at early times.
Hence, the H abundance is higher than that of H$_{2}$ in the slow warm-up model, while the H$_{2}$ abundance is higher in the fast warm-up model.
In the slow warm-up model, the H$^{+}$ ion is efficiently formed by the ionization of H atom by cosmic rays. 
On the other hand, in the fast warm-up model, the H$^{+}$ ion is formed by the reaction between H$_{2}$ and cosmic rays, leading to H, H$^{+}$, and electrons, which is slower than the ionization of H atoms by cosmic rays.
Hence, the formation of the H$^{+}$ ion is more efficient in the slow warm-up model. 
Cosmic rays also produce H$_{2}^{+}$ from H$_{2}$ ionization.  
The H$_{2}^{+}$ ion subsequently reacts with H$_{2}$ forming the H$_{3}^{+}$ ion in the fast warm-up model, whereas H$_{2}^{+}$ reacts with H to form H$_{2}$ and H$^{+}$ in the slow warm-up model.
Therefore, in the slow warm-up model, the H$^{+}$ abundance is larger than that of H$_{3}^{+}$, while the opposite is true in the fast warm-up model.
 
The H$^{+}$ ion contributes to destruction of HC$_{5}$N to form HC$_{5}$N$^{+}$ in the slow warm-up model. 
The reaction between HC$_{5}$N$^{+}$ and electrons cannot produce HC$_{5}$N.
In the fast warm-up model, on the other hand, the reactions with H$_{3}^{+}$ and HCO$^{+}$ destroy HC$_{5}$N, and both reactions form H$_{2}$C$_{5}$N$^{+}$, which reacts with electron and partially goes back to HC$_{5}$N. 
Reaction (\ref{equ:r2}) continues to form HC$_{5}$N in both models and this reaction is particularly important for the slow warm-up model. 
The C$_{4}$H$_{2}$ molecules are formed from large hydrocarbons (e.g., C$_{4}$H) and their ions (e.g., C$_{4}$H$_{3}^{+}$) in both models, not bottom-up formation starting from CH$_{4}$. 
Therefore, the carbon-chain formation starting from CH$_{4}$ is no longer an important formation mechanism of cyanopolyynes for the highest cosmic-ray ionization models.

In summary, the lower limit of HC$_{5}$N observed in the G28.28--0.36 MYSO can be reproduced with temperatures above $\sim 115$ K in all of the models with the typical cosmic-ray ionization rate of $1.3 \times 10^{-17}$ s$^{-1}$.
In the case of the Slow warm-up timescale, the models with high cosmic-ray ionization rates cannot reproduce the observed abundance of HC$_{5}$N in G28.28--0.36.
On the other hand, the model with Fast warm-up timescale and the highest cosmic-ray ionization rate can maintain the high abundance of HC$_{5}$N.

\subsection{Comparisons among Carbon-Chain Species} \label{sec:dis3}

\begin{figure}
\figurenum{9}
 \begin{center}
  \includegraphics[bb=0 20 540 270, scale=0.65]{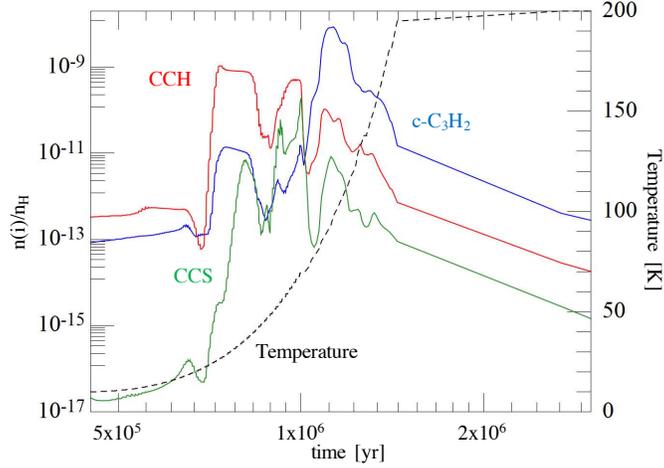}
 \end{center}
\caption{Abundances of carbon-chain species in the gas phase during the warm-up period (Model M3). The red, green, and blue curves indicate results for CCH, CCS, and $c$-C$_{3}$H$_{2}$, respectively. The black dashed line indicates temperature. \label{fig:f5}}
\end{figure}

We found unusual features of the cyanopolyynes during the warm-up period; in particular, they are formed initially in the gas phase, subsequently  accumulate in the bulk of the icy mantle, and then desorb into the gas phase, with their gas-phase peak abundances attained when they desorb.
The carbon-chain molecule CH$_{3}$CCH shows a similar feature (Section \ref{sec:dis2}).

Figure \ref{fig:f5} shows the abundances of CCH (red), CCS (green), and $c$-C$_{3}$H$_{2}$ (blue) during the warm-up period.
The species CCH and CCS show abundance peaks just after the temperature rises to 25 K.
The CCH molecule is efficiently destroyed by the reaction with atomic oxygen (O) during $1.0 \times 10^{6} \leq t \leq 1.07 \times 10^{6}$ yr, when the temperature rises from $\approx 70$ K to $\approx 80$ K.
The subsequent decrease in the gas-phase CCH abundance is caused by its efficient destruction by reaction with H$_{2}$, which occurs efficiently with the temperature of $\geq 90$ K, at a time of $t \geq 1.1 \times 10^{6}$ yr \citep{2011ApJ...743..182H}.
Dicarbon sulfide, or CCS, is destroyed by the reaction with atomic oxygen (O). 
The atomic oxygen desorbs at temperatures above 60 K \citep{2015ApJ...801..120H} at a time of $t \geq 1.0 \times 10^{6}$ yr.
Because of these rapid gas-phase reactions, desorption from dust surfaces is not important for the synthesis of gas-phase CCH and CCS.

On the other hand, $c$-C$_{3}$H$_{2}$ shows its gas-phase peak abundance after its direct evaporation from the dust surface at temperatures above 90 K at a time of $t \geq 1.1 \times 10^{6}$ yr, as is the case for the  cyanopolyynes and CH$_{3}$CCH.
The cyclic species $c$-C$_{3}$H$_{2}$ is produced in the gas-phase both in the pre-warm-up period and during the warm-up period starting with the sublimation of CH$_{4}$.  
The main formation pathway of $c$-C$_{3}$H$_{2}$ is the dissociative recombination reaction of $c$ - C$_{3}$H$_{3}^{+}$:
\begin{equation} \label{rea:c-C3H21}
c-{\rm {C}}_{3}{\rm {H}}_{3}^{+} + {\rm {e}}^{-} \rightarrow c-{\rm {C}}_{3}{\rm {H}}_{2} + {\rm {H}}.
\end{equation}
Gaseous $c$-C$_{3}$H$_{2}$, formed in the low-temperature portion of the warm-up period, is subsequently adsorbed onto dust grains, from which it desorbs after $\sim 1.0 \times 10^{6}$ yr at temperatures $T \geq 65$ K.
It is efficiently destroyed by the reaction with H$_{2}$ at temperatures above 120 K ($t \geq 1.2 \times 10^{6}$ yr).
Hence, the efficient destruction of $c$-C$_{3}$H$_{2}$ occurs after it desorbs.
All species in Figure \ref{fig:f5} decrease in abundance with parallel curves after $1.5 \times 10^{6}$ yr.

In summary, there are largely two types of carbon-chain species in warm-up regions.
One type shows peak abundances just after their desorption from dust grains; these include cyanopolyynes, CH$_{3}$CCH, and $c$-C$_{3}$H$_{2}$.
The other type shows peak abundances just after the temperature rises to 25 K (CCH and CCS).
The latter include relatively reactive species, which are easily destroyed in the gas phase.
On the other hand, cyanopolyynes are not destroyed by reactions with H$_{2}$ and O, but react with ions such as H$^{+}$, H$_{3}^{+}$, and HCO$^{+}$, and so eventually decrease.  

Such differences among carbon-chain species seem to explain the observational results in HMPOs \citep{2018ApJ...854..133T, 2019ApJ...872..154T} and massive cluster-forming regions \citep{2018ApJ...855...45S}.
\citet{2018ApJ...854..133T} suggested that HC$_{3}$N is newly formed in the warm and well-shielded dense gas around HMPOs, and \citet{2019ApJ...872..154T} reported that the detection rates  for the cyanopolyynes, defined by the number of sources where target molecules were detected divided by the number of total observed sources $\times 100$,  are higher around HMPOs than  in low-mass protostars, but that of CCS is lower.
\citet{2018ApJ...855...45S} concluded that HC$_{3}$N tends to be abundant, whereas  CCS detection is rare in massive cluster-forming clumps, where it has  probably been already destroyed by reaction with O because the dust temperature should be higher than the sublimation temperature of atomic oxygen.  On the other hand, HC$_{3}$N can still be abundant in such high temperature regions.

\subsection{Implication for Chemical Diversity around Massive Young Stellar Objects (MYSOs)} \label{sec:dis4}

As  already mentioned in Section \ref{sec:intro}, the different timescales of the starless core phase \citep{2013ChRv..113.8981S} and the penetration of the interstellar radiation field \citep{2016A&A...592L..11S} were proposed as possible origins of chemical diversity in low-mass star-forming regions.
High-mass stars are usually born in giant molecular clouds (GMCs), and there is a possibility that the formation of high-mass starless cores starts long after the clouds were well shielded against the interstellar radiation field unless GMCs have clumpy structures.
In the former case, there should be enough time for atomic carbon to convert into CO molecules.
In particular, the interstellar radiation field may be important for the outer edge of GMCs but not for the inner regions, as is the well-known case of photon-dominated regions (PDRs).

Chemical diversity can also be caused by the dependence of abundance on the heating rate.  
For example, CH$_{3}$OH shows higher peak abundances in those models with a Fast heating timescale.
On the other hand, as discussed above, HC$_{5}$N tends to be more abundant in the models with the Slow heating timescale because there is enough time for HC$_{5}$N to accumulate in the bulk of the icy mantle before desorption.
This may be one possible origin of the chemical diversity around MYSOs \citep{2017ApJ...844...68T,2018ApJ...866..150T}. 

The heating timescale of warm-up regions depends not only on stellar masses but also on the relationship between the size of the warming region and the infall velocity \citep{2008ApJ...674..984A} as given by
\begin{equation} \label{equ:th}
t_{\rm {h}} \propto \frac{R_{\rm {warm}}}{V_{\rm {infall}}},
\end{equation}
where $t_{\rm {h}}$, $R_{\rm {warm}}$, and $V_{\rm {infall}}$ are the heating timescale, the size of the warm region, and the infall velocity, respectively.
If $R_{\rm {warm}}$ becomes larger or $V_{\rm {infall}}$ becomes smaller, $t_{\rm {h}}$ will be longer.
Such a condition will lead to HC$_{5}$N-rich/CH$_{3}$OH-poor envelopes around MYSOs, which is similar to the case of G28.28$-$0.36.

The size of warm regions and their infall velocity relate to other physical conditions.
For example, the size of warm regions depends upon  the luminosity of the central stars, and the density structure \citep{2004A&A...414..409N}, while
the infall velocity is related to the density, rotating motion, magnetic field, and radiation pressure \citep[e.g.,][]{2014A&A...562A..82S}.
Consequently, the heating timescale is also related to the physical conditions in star-forming regions.
Observations investigating relationships between the chemical diversity and Equation (\ref{equ:th}) are necessary to constrain the proposed scenario.

The possibility of  UV radiation to cause chemical diversity should be considered in cluster regions.
The first born star should affect its environment and the strong UV radiation it emits can destroy CO molecules to form C and/or C$^{+}$, which will be precursors of carbon-chain species.  In that case, we would expect chemical diversity among sources in the same cluster region.
As discussed in Section \ref{sec:dis5}, the cosmic-ray ionization rate significantly affects the abundances of cyanopolyynes. 
Cosmic rays can penetrate deeper dense regions compared to the UV radiation. 
Hence, the effects of cosmic rays may be essential particular in the dense cores including hot cores.

The binding energies are affected by significant errors (Table \ref{tab:bind} in Appendix \ref{sec:bind}) and they change with ice-mantle surface composition \citep[e.g.,][]{2018A&A...619A.111N}. 
More laboratory work should be dedicated to binding energy measurements; meanwhile, detailed comparison between model predictions and observational results toward star-forming regions will help to put constraints on these important parameters.

\section{Conclusions} \label{sec:con}

We have investigated cyanopolyyne chemistry around MYSOs with warm-up hot-core models using the Nautilus code, and motivated by recent observational results toward MYSOs \citep{2017ApJ...844...68T,2018ApJ...866..150T}.
The main conclusions of this paper are as follows: 
\begin{enumerate}
\item Cyanopolyynes are formed via  neutral-neutral and ion-neutral reactions in the lukewarm gas ($25 < T < 100 $ K).
The sublimation of CH$_{4}$ and C$_{2n}$H$_{2}$ from dust grains enhances key reactions for the formation of cyanopolyynes. 
They are simultaneously accumulated on the dust surface and in the bulk of icy mantles in this temperature range.
Cyanopolyynes sublimate into the gas phase when the temperature rises above $\sim 100$ K and reach their peak abundances in the gas phase.
\item The carbon chain CH$_{3}$CCH shows similar characteristics to cyanopolyynes. It is formed from CH$_{4}$ sublimated from dust grains and accumulates onto dust grains before the temperature reaches its sublimation temperature of 70 K. 
After the temperature reaches 70 K, CH$_{3}$CCH desorbs from dust grains and shows the gas-phase peak abundance.
\item Models with a longer warm-up period enable these species to accumulate to a greater extent in the bulk of icy mantles.
\item Our model results can reproduce the lower limits of HC$_{5}$N, CH$_{3}$CCH, and CH$_{3}$OH observed in the G28.28$-$0.36 MYSO, where HC$_{5}$N is particularly abundant \citep{2017ApJ...844...68T}.
Their observed abundances are reproduced best just after their sublimation from dust grains.
\item The species CCH and CCS show their peak abundances just after they are formed in the gas phase triggered by the evaporation of CH$_{4}$. 
The chemistry that produces them is the WCCC chemistry.
They are reactive species and destroyed by H$_{2}$ or O in the gas phase.
On the other hand, cyanopolyynes, CH$_{3}$CCH, and $c$-C$_{3}$H$_{2}$ are relatively stable species and can accumulate in the bulk of icy mantles before destruction in the gas phase.
This implies that there are largely two types of carbon-chain species.
Such results seem to explain the higher detection rates of cyanopolyynes than that of CCS in HMPOs and cluster-forming regions.
\item HC$_{5}$N-rich and CH$_{3}$OH-poor envelopes around MYSOs may reflect different heating timescales.
The heating timescale depends not only on the mass of central stars but also on the relationships between the size of warm regions and their infall velocity, which in turn relate to various physical conditions, including  luminosity, density, magnetic field, rotating motion, and radiation pressure.
\end{enumerate}

\acknowledgments

K. T. would like to thank the University of Virginia for providing the funds for her postdoctoral fellowship in the Virginia Initiative on Cosmic Origins (VICO) research program. E. H. is grateful for support from the National Science Foundation through grant 1514844.
%

\vspace{5mm}


\software{Nautilus \citep{2016MNRAS.459.3756R}}



\appendix

\section{Binding Energy} \label{sec:bind}

Table \ref{tab:bind} summarizes binding energies of major species and key species in this paper utilized in our model.
These binding energy values and errors are taken from the KIDA database \citep{2017MolAs...6...22W}.

\floattable
\begin{deluxetable}{cccc}[h]
\tabletypesize{\scriptsize}
\tablecaption{Binding energies of major species \label{tab:bind}}
\tablewidth{0pt}
\tablehead{
\colhead{Species} & \colhead{Binding Energy (K)} & \colhead{Species} & \colhead{Binding Energy (K)}
}
\startdata
H & $650 \pm 195$ & C$_{2}$H$_{2}$ & $2587 \pm 776$ \\
H$_{2}$ & $440 \pm 132$ & C$_{4}$H$_{2}$ & 4187 \\
C & 4000 & C$_{6}$H$_{2}$ & 5787 \\
O & $1660 \pm 60$ & HC$_{3}$N & 4580 \\
N & 800 & HC$_{5}$N & 6180 \\
CO & 1150 & HC$_{7}$N & 7780 \\
CN & 1600 & CH$_{3}$OH & 5534 \\
CH$_{4}$ & 1300 & CH$_{3}$CCH & 4287 \\
\enddata
\tablecomments{These values are applied for species on amorphous water ice surface \citep{2017MolAs...6...22W}.}
\end{deluxetable}	


\begin{thebibliography}{}
\bibitem[Acharyya \& Herbst(2017)]{2017ApJ...850..105A} Acharyya, K., \& Herbst, E.\ 2017, \apj, 850, 105 
\bibitem[Aikawa et al.(2008)]{2008ApJ...674..984A} Aikawa, Y., Wakelam, V., Garrod, R.~T., \& Herbst, E.\ 2008, \apj, 674, 984
\bibitem[Balucani et al.(2015)]{2015MNRAS.449L..16B} Balucani, N., Ceccarelli, C., \& Taquet, V.\ 2015, \mnras, 449, L16
\bibitem[Belloche et al.(2013)]{2013A&A...559A..47B} Belloche, A., M{\"u}ller, H.~S.~P., Menten, K.~M., Schilke, P., \& Comito, C.\ 2013, \aap, 559, A47
\bibitem[Benson et al.(1998)]{1998ApJ...506..743B} Benson, P.~J., Caselli, P., \& Myers, P.~C.\ 1998, \apj, 506, 743
\bibitem[Bergantini et al.(2018)]{2018ApJ...852...70B} Bergantini, A., G{\'o}bi, S., Abplanalp, M.~J., \& Kaiser, R.~I.\ 2018, \apj, 852, 70
\bibitem[Beuther et al.(2002)]{2002ApJ...566..945B} Beuther, H., Schilke, P., Menten, K.~M., et al.\ 2002, \apj, 566, 945
\bibitem[Bonfand et al.(2017)]{2017A&A...604A..60B} Bonfand, M., Belloche, A., Menten, K.~M., Garrod, R.~T., \& M{\"u}ller, H.~S.~P.\ 2017, \aap, 604, A60
\bibitem[Burkhardt et al.(2018)]{2018MNRAS.474.5068B} Burkhardt, A.~M., Herbst, E., Kalenskii, S.~V., et al.\ 2018, \mnras, 474, 5068 
\bibitem[Caselli \& Ceccarelli(2012)]{2012A&ARv..20...56C} Caselli, P., \& Ceccarelli, C.\ 2012, \aapr, 20, 56
\bibitem[Esplugues et al.(2013)]{2013A&A...559A..51E} Esplugues, G.~B., Cernicharo, J., Viti, S., et al.\ 2013, \aap, 559, A51
\bibitem[Favre et al.(2018)]{2018ApJ...859..136F} Favre, C., Ceccarelli, C., L{\'o}pez-Sepulcre, A., et al.\ 2018, \apj, 859, 136 
\bibitem[Fontani et al.(2017)]{2017A&A...605A..57F} Fontani, F., Ceccarelli, C., Favre, C., et al.\ 2017, \aap, 605, A57 
\bibitem[Garrod(2013)]{2013ApJ...765...60G} Garrod, R.~T.\ 2013, \apj, 765, 60
\bibitem[Garrod \& Herbst(2006)]{2006A&A...457..927G} Garrod, R.~T., \& Herbst, E.\ 2006, \aap, 457, 927 
\bibitem[Green et al.(2014)]{2014MNRAS.443.2252G} Green, C.-E., Green, J.~A., Burton, M.~G., et al.\ 2014, \mnras, 443, 2252
\bibitem[Hassel et al.(2011)]{2011ApJ...743..182H} Hassel, G.~E., Harada, N., \& Herbst, E.\ 2011, \apj, 743, 182 
\bibitem[Hassel et al.(2008)]{2008ApJ...681.1385H} Hassel, G.~E., Herbst, E., \& Garrod, R.~T.\ 2008, \apj, 681, 1385 
\bibitem[He et al.(2015)]{2015ApJ...801..120H} He, J., Shi, J., Hopkins, T., et al.\ 2015, \apj, 801, 120
\bibitem[Herbst \& Millar(2008)]{HM2008} Herbst E. \& Millar, T.~J. 2008, in Low Temperatures and Cold Molecules,  Smith, I. W. M. ed. (London: Imperial College Press), p. 1
\bibitem[Herbst \& van Dishoeck(2009)]{2009ARA&A..47..427H} Herbst, E., \& van Dishoeck, E.~F.\ 2009, \araa, 47, 427 
\bibitem[Hirota et al.(2009)]{2009ApJ...699..585H} Hirota, T., Ohishi, M., \& Yamamoto, S.\ 2009, \apj, 699, 585
\bibitem[Hudson \& Moore(2018)]{2018ApJ...857...89H} Hudson, R.~L., \& Moore, M.~H.\ 2018, \apj, 857, 89
\bibitem[Jaber Al-Edhari et al.(2017)]{2017A&A...597A..40J} Jaber Al-Edhari, A., Ceccarelli, C., Kahane, C., et al.\ 2017, \aap, 597, A40 
\bibitem[J{\o}rgensen et al.(2002)]{2002AA...389..908J} J{\o}rgensen, J.~K., Sch{\"o}ier, F.~L., \& van Dishoeck, E.~F.\ 2002, \aap, 389, 908
\bibitem[Lee et al.(1996)]{1996A&A...311..690L} Lee, H.-H., Herbst, E., Pineau des Forets, G., Roueff, E., \& Le Bourlot, J.\ 1996, \aap, 311, 690 
\bibitem[Li et al.(2013)]{2013A&A...555A..14L} Li, X., Heays, A.~N., Visser, R., et al.\ 2013, \aap, 555, A14 
\bibitem[McGuire et al.(2018)]{2018Sci...359..202M} McGuire, B.~A., Burkhardt, A.~M., Kalenskii, S., et al.\ 2018, Science, 359, 202 
\bibitem[Nguyen et al.(2018)]{2018A&A...619A.111N} Nguyen, T., Baouche, S., Congiu, E., et al.\ 2018, \aap, 619, A111
\bibitem[Nomura \& Millar(2004)]{2004A&A...414..409N} Nomura, H., \& Millar, T.~J.\ 2004, \aap, 414, 409 
\bibitem[Padovani et al.(2016)]{2016A&A...590A...8P} Padovani, M., Marcowith, A., Hennebelle, P., \& Ferri{\`e}re, K.\ 2016, \aap, 590, A8
\bibitem[Reboussin et al.(2014)]{2014MNRAS.440.3557R} Reboussin, L., Wakelam, V., Guilloteau, S., \& Hersant, F.\ 2014, \mnras, 440, 3557 
\bibitem[Ruaud et al.(2016)]{2016MNRAS.459.3756R} Ruaud, M., Wakelam, V., \& Hersant, F.\ 2016, \mnras, 459, 3756   
\bibitem[Sakai et al.(2009)]{2009ApJ...697..769S} Sakai, N., Sakai, T., Hirota, T., Burton, M., \& Yamamoto, S.\ 2009, \apj, 697, 769
\bibitem[Sakai et al.(2008)]{2008ApJ...672..371S} Sakai, N., Sakai, T., Hirota, T., \& Yamamoto, S.\ 2008, \apj, 672, 371
\bibitem[Sakai \& Yamamoto(2013)]{2013ChRv..113.8981S} Sakai, N., \& Yamamoto, S.\ 2013, Chemical Reviews, 113, 8981 
\bibitem[Shimoikura et al.(2018)]{2018ApJ...855...45S} Shimoikura, T., Dobashi, K., Nakamura, F., Matsumoto, T., \& Hirota, T.\ 2018, \apj, 855, 45 
\bibitem[Shingledecker \& Herbst(2018)]{2018PCCP...20.5359S} Shingledecker, C.~N., \& Herbst, E.\ 2018, Physical Chemistry Chemical Physics (Incorporating Faraday Transactions), 20, 5359 
\bibitem[Skouteris et al.(2018)]{2018ApJ...854..135S} Skouteris, D., Balucani, N., Ceccarelli, C., et al.\ 2018, \apj, 854, 135
\bibitem[Spezzano et al.(2016)]{2016A&A...592L..11S} Spezzano, S., Bizzocchi, L., Caselli, P., Harju, J., \& Br{\"u}nken, S.\ 2016, \aap, 592, L11
\bibitem[Sugiyama et al.(2014)]{2014A&A...562A..82S} Sugiyama, K., Fujisawa, K., Doi, A., et al.\ 2014, \aap, 562, A82 
\bibitem[Suzuki et al.(1992)]{1992ApJ...392..551S} Suzuki, H., Yamamoto, S., Ohishi, M., et al.\ 1992, \apj, 392, 551
\bibitem[Taniguchi et al.(2018a)]{2018ApJ...866...32T} Taniguchi, K., Miyamoto, Y., Saito, M., et al.\ 2018a, \apj, 866, 32 
\bibitem[Taniguchi et al.(2016a)]{2016ApJ...817..147T} Taniguchi, K., Ozeki, H., Saito, M., et al.\ 2016a, \apj, 817, 147 
\bibitem[Taniguchi \& Saito(2017)]{2017PASJ...69L...7T} Taniguchi, K., \& Saito, M.\ 2017, \pasj, 69, L7
\bibitem[Taniguchi et al.(2017)]{2017ApJ...844...68T} Taniguchi, K., Saito, M., Hirota, T., et al.\ 2017, \apj, 844, 68 
\bibitem[Taniguchi et al.(2018b)]{2018ApJ...866..150T} Taniguchi, K., Saito, M., Majumdar, L., et al.\ 2018b, \apj, 866, 150 
\bibitem[Taniguchi et al.(2016b)]{2016ApJ...830..106T} Taniguchi, K., Saito, M., \& Ozeki, H.\ 2016b, \apj, 830, 106 
\bibitem[Taniguchi et al.(2018c)]{2018ApJ...854..133T} Taniguchi, K., Saito, M., Sridharan, T.~K., \& Minamidani, T.\ 2018c, \apj, 854, 133 
\bibitem[Taniguchi et al.(2019)]{2019ApJ...872..154T} Taniguchi, K., Saito, M., Sridharan, T.~K., \& Minamidani, T.\ 2019, \apj, 872, 154 
\bibitem[van Dishoeck(2018)]{2018IAUS..332....3V} van Dishoeck, E.~F.\ 2018, IAU Symposium, 332, 3
\bibitem[Visser et al.(2009)]{2009A&A...503..323V} Visser, R., van Dishoeck, E.~F., \& Black, J.~H.\ 2009, \aap, 503, 323 
\bibitem[Viti et al.(2004)]{2004MNRAS.354.1141V} Viti, S., Collings, M.~P., Dever, J.~W., McCoustra, M.~R.~S., \& Williams, D.~A.\ 2004, \mnras, 354, 1141 
\bibitem[Wakelam et al.(2017)]{2017MolAs...6...22W} Wakelam, V., Loison, J.-C., Mereau, R., \& Ruaud, M.\ 2017, Molecular Astrophysics, 6, 22
\end{thebibliography}
\end{document}